\documentclass[lettersize,journal]{IEEEtran}
\usepackage{amsmath,amsfonts}
\usepackage{algorithmic}
\usepackage{algorithm}
\usepackage{array}
\usepackage{enumerate}
\usepackage[caption=false,font=normalsize,labelfont=sf,textfont=sf]{subfig}
\usepackage{textcomp}
\usepackage{stfloats}
\usepackage{url}
\usepackage{verbatim}
\usepackage{graphicx}
\usepackage{subfig}
\usepackage{cite}
\usepackage{balance}
\usepackage[numbers,sort&compress]{natbib}
\usepackage{lipsum}
\usepackage{textcomp,mathcomp}
\usepackage{xcolor}
\usepackage{multirow}

\begin{document}

\title{A Novel Modular, Reconfigurable Battery Energy Storage System: Design, Control, and Experimentation}

\author{Amir Farakhor,~\IEEEmembership{Student Member,~IEEE}, Di Wu,~\IEEEmembership{Senior Member,~IEEE},
		Yebin Wang,~\IEEEmembership{Senior Member,~IEEE}, and Huazhen Fang,~\IEEEmembership{Member,~IEEE}

\thanks{This work was supported in part by the U.S. National Science Foundation under Awards CMMI-1763093 and CMMI-1847651, and in part by the U.S. Department of Energy, Office of Electricity through the Energy Storage Program.}

\thanks{A. Farakhor and H. Fang (corresponding author) are with the Department of Mechanical Engineering, University of Kansas, Lawrence, KS, USA (Email: fang@ku.edu, a.farakhor@ku.edu).} 
\thanks{D. Wu is with the Pacific Northwest National Laboratory, Richland, WA, USA (Email: di.wu@pnnl.gov).}
\thanks{Y. Wang is with the Mitsubishi Electric Research Laboratories, Cambridge, MA, USA (Email: yebinwang@merl.com).}
}

\markboth{}
{Shell \MakeLowercase{\textit{et al.}}: A Sample Article Using IEEEtran.cls for IEEE Journals}

\maketitle

\begin{abstract}
This paper presents a novel modular, reconfigurable battery energy storage system. The proposed design is characterized by a tight integration of reconfigurable power switches and DC/DC converters. This characteristic enables isolation of faulty cells from the system and allows fine power control for individual cells toward optimal system-level performance. An optimal power management approach is developed to extensively exploit the merits of the proposed design. Based on receding-horizon convex optimization, this approach aims to minimize the total power losses in charging/discharging while allocating the power in line with each cell's condition to achieve state-of-charge (SoC) and temperature balancing. By appropriate design, the approach manages to regulate the power of a cell across its full SoC range and guarantees the feasibility of the optimization problem. We perform extensive simulations and further develop a lab-scale prototype to validate the proposed system design and power management approach. 
\end{abstract}


\section*{Nomenclature}
\addcontentsline{toc}{section}{Nomenclature}
\begin{IEEEdescription}[\IEEEusemathlabelsep\IEEEsetlabelwidth{$V_1,V_2,V_3$}]

\item[\textbf{Variables}]

\item[$S$] Reconfiguration switches--binary (1/0) variable
\item[$V_t^*$] RBESS reference output voltage
\item[$V_C^{\textrm{max}}$] Maximum output voltage of DC/DC converters
\item[$I_C^{\textrm{max}}$] Maximum output current of DC/DC converters
\item[$v$] Cell voltage
\item[$u$] Cell open-circuit voltage
\item[$i_L$] Cell current
\item[$P_b$] Cell internal power
\item[$P$] Cell output power
\item[$E$] Cell energy
\item[$P_l$] Module power losses
\item[$J$] Total power losses
\item[$P_{\textrm{out}}$] RBESS output power
\item[$I_{\textrm{out}}$] RBESS output charging/discharging current
\item[$\dot{Q}_{\textrm{cnd}}$] Conductive heat transfer rate
\item[$\dot{Q}_{\textrm{conv}}$] Convective heat transfer rate
\item[$\xi$] Slack variable
\item[$z$] Optimization variables

\vspace{0.25cm}
\item[\textbf{Parameters}]

\item[$n$] Number of battery cells
\item[$n_s$] Number of cells in series
\item[$n_p$] Number of cells in parallel
\item[$\cal{J}$] Set of in-service cells
\item[$L$] Inductor for DC/DC converters
\item[$C$] Output capacitor for DC/DC converters
\item[$\bar{Q}$] Cell capacity
\item[$q, q^{\textrm{min}}, q^{\textrm{max}}$] SoC of a cell, and its lower and upper limits
\item[$q_{\textrm{avg}}$] Average SoC
\item[$\Delta q$] SoC imbalance tolerance
\item[$\Delta E$] Energy imbalance tolerance
\item[$\alpha$] Intercept coefficient of the SoC/OCV line segment
\item[$\beta$] Slope coefficient of the SoC/OCV line segment
\item[$R$] Cell internal resistance
\item[$R_C$] Resistance to capture the power losses of DC/DC converters
\item[$R_{\textrm{cnd}}$] Conductive thermal resistance
\item[$R_{\textrm{conv}}$] Convective thermal resistance
\item[$C_{\textrm{th}}$] Thermal capacitance
\item[$T, T^{\textrm{max}}$] Cell temperature and its upper limit
\item[$T_{\textrm{avg}}$] Average temperature
\item[$T_{\textrm{env}}$] Environmental temperature
\item[$\Delta T$] Temperature imbalance tolerance
\item[$\lambda$] Penalty weight for the multi-objective optimization
\item[$\Delta t$] Sampling time
\item[$H$] Optimization horizon
\end{IEEEdescription}

\section{Introduction}
\IEEEPARstart{L}{ithium-ion} battery energy storage systems (BESSs) have proven themselves as an enabling technology for various applications, including electric cars, electric aircraft, smart grid, and space systems\cite{1512466,bills2020universal,ROSEWATER2015460,9353049}. Despite their high energy density and long cycle life, lithium-ion batteries suffer from safety risks, which trace to the high reactivity of lithium and flammability of the commonly used electrolyte solutions and are exacerbated by side reactions, aging, and degradation \cite{FENG2018246}. Hence, it is imperative to ensure their safe and reliable operation, particularly in safety-critical applications \cite{7433464}. Reconfigurable BESS (RBESS) have attracted much attention as a promising means to achieve this end. An RBESS characteristically uses power electronics switches to make the connection among the constituent cells reconfigurable, providing the capability to bypass faulty cells without interrupting the operation of the system \cite{7442763}. This feature overcomes the vulnerability of conventional hardwired BESS to single-cell failures due to the fixed configuration \cite{9211456}. This paper proposes a novel RBESS design that integrates reconfigurable switches with DC/DC converters. The new design uses the switches to reconfigure the connectivity of the cells for the sake of safety and meanwhile, leverages the converters to achieve robust power management and supply from cell to system level. Further, we present an optimal power management algorithm and develop a lab-scale prototype to validate the proposed RBESS design and control.

\subsection{Literature Review}

This paper centers around the RBESS circuit design and power management. Therefore, we survey the literature on the two dimensions one by one.  The review will also encompass some recent studies about hardwired BESS due to the relevance. 

\subsubsection{Review of RBESS circuit design}

The literature has presented two main ways to design RBESS circuit architectures. The first one builds and integrates a circuit of controllable power electronics switches with the cells. By controlling the switches, a cell can be put into or cut off from the connection with other cells when a fault occurs. The circuit topology plays a key role in the level of reconfigurability, functional flexibility, and circuit complexity. The study in \cite{5744772} shows a switching circuit for a series-connected battery pack, which uses only two switches for every cell. More sophisticated topologies can provide more versatile reconfiguration, though at the expense of using larger numbers of switches. The work in \cite{6126056} considers a string of modules of parallel-connected cells in series and allows switch-based bypass of any cell or module.  The circuit topologies proposed in \cite{6165857,4840570} place five and six switches, respectively, around each cell to realize arbitrary series and parallel connection among the cells, and the one shown in \cite{9720183} uses only four to achieve the same end. 
It is interesting to point out that  reconfigurable switching circuits can  enable more functions than just isolating faulty cells. A battery system designed in \cite{8099064} dynamically connects some battery modules in series or parallel to produce multi-level and even AC voltage output. In \cite{9090983}, an inverter based on switching circuits combines batteries and supercapacitors for hybrid energy storage. While   easily reconfigurable, switching circuits are unable to do cell-level charging/discharging power regulation. This limitation will lead to unbalanced use of the cells following the reconfiguration-based bypass of a cell. 

Another important way for RBESS design uses converters. A converter can not only connect a cell into the circuit, but also charge or discharge it at a controlled current, voltage, or power. This capability offers a higher degree of freedom for cell-level control to enhance the balanced use of the cells.   In \cite{7116565}, a centralized converter interfaces with a few cells, and it contains a selector to select and put the cells into operation. The work in \cite{7378996} pairs converters, which by design include power switches to offer a bypass mode, with individual cells one by one to form a reconfigurable pack. However, converter-based RBESS architectures are unable to offer flexible topology changes, compared to the switching circuits. As such, only series-connected RBESS is considered in \cite{7116565,7378996}. 
Converters have also found their way into hardwired BESS.  The studies in \cite{7544629,8768008,9520124} integrate DC/DC converters with the cells to enable   cell balancing   and power loss minimization. However, the fixed hardwired connections among the cells make them susceptible to single-cell faults. 

\subsubsection{Review of RBESS control and power management}

For a given switching-circuit-based RBESS, an essential question is how to control its topology. A main approach lies in finding out a connection topology to meet load requirements. For example, a heuristic method in \cite{6126056} groups battery cells into modules and chooses a minimum set of modules on a load request to perform charging/discharging. The method of dynamic programming finds use in identifying an optimal configuration topology in \cite{6165857}, and rule-based bypassing mechanisms are proposed in \cite{4840570} to control the switches for stable or responsive voltage supply. The study in \cite{9720183} exploits the switching circuit to achieve cell equalization through a hierarchical strategy that combines intra-module equalization and system-level reconfiguration. Considering the switching circuits in~\cite{5744772,6126056}, the study in \cite{7782856} proposes to sort   cells   according to their capacity and then reconfigure them  into serial strings to restrain the cell imbalance. 

On a related note, the literature has included a few studies about power management for hardwired BESS based on converters. For converter-based RBESS, the main subject of investigation is developing control approaches to manage the cell-to-system-level operation. Optimal control   has shown as a useful solution in this regard. However, due to a lithium-ion battery's nonlinear characteristics, optimization problems posed for the power management are often non-convex, and thus defy the computation of global optima. To mitigate the issue, the studies in~\cite{7544629,8768008} choose to leverage a   battery model convexification approach proposed in \cite{MURGOVSKI201292} to formulate convex optimal power control problems. The convexification therein involves a linear approximation of the relationship between the state-of-charge (SoC) and the open-circuit-voltage (OCV), thus restricting the proposed methods to only limited operating ranges. It is also possible to use a linear battery model to achieve optimal control that even admits closed-form solutions \cite{7518611}, but the method is also just applicable to operating ranges in which the linear model is accurate. 

As another valuable approach, distributed power management treats the cells as independent agents and makes them perform individual control toward a global goal. One can decompose a global optimization task and distribute it among individual cells for charging or load sharing, as shown in \cite{9520124}. As an alternative way, distributed consensus control is applied in \cite{7995102} to achieve SoC balancing among the cells. Not requiring numerical optimization, this method offers fast computation. But the lack of optimality makes it in need of more time to converge.  

\begin{table*}[t]
	\renewcommand{\arraystretch}{1.2}
	\caption{Characteristics of different BESS structures.}
	\centering
	\label{TABLE_1}
  \begin{tabular}{>{\color{black}}l>{\color{black}}l>{\color{black}}l>{\color{black}}l>{\color{black}}l>{\color{black}}l>{\color{black}}l>{\color{black}}l|>{\color{black}}l>{\color{black}}l|>{\color{black}}l>{\color{black}}l>{\color{black}}l|>{\color{black}}cl}
			\hline\hline \\[-3.72mm]
&\multicolumn{7}{c|}{\parbox[s]{2.5cm}{\centering Switch-based RBESS}}&\multicolumn{2}{c|}{\parbox[t]{1.6cm}{\centering Converter-based RBESS}}&\multicolumn{3}{c|}{\parbox[t]{2cm}{Converter-based hardwired BESS}}&\multirow{2}{*}{Proposed design}\\[+2.5mm] 
			&\cite{5744772}&\cite{6126056}&\cite{6165857}&\cite{4840570}&\cite{9720183}&\cite{8099064}&\cite{9090983}&\hspace{3pt}\cite{7116565}&\cite{7378996}&\cite{7544629}&\cite{8768008}&\cite{9520124}\\[+0.5mm] \hline 
			Safety (Bypass of faulty cells) 		& \checkmark & \checkmark & \checkmark & \checkmark & \checkmark & \checkmark & \checkmark & \hspace{7pt}\checkmark & \hspace{4pt}\checkmark & \hspace{3pt}$\times$	& \hspace{3pt}$\times$	& $\hspace{3pt}\times$	& \checkmark \\
			Output voltage regulation  			& $\times$ 	& $\times$	& $\times$	& $\times$	& $\times$	& $\times$	& $\times$	&\hspace{4pt} \checkmark & \hspace{4pt}\checkmark & \hspace{3pt}$\times$ 	& \hspace{3pt}$\times$ 	& \hspace{3pt}$\times$	& \checkmark \\
			SoC balancing 						& \checkmark & \checkmark & \checkmark & \checkmark & \checkmark & $\times$ 	& \checkmark & \hspace{7pt}\checkmark & \hspace{4pt}\checkmark & \hspace{3pt}\checkmark & \hspace{3pt}\checkmark & \hspace{3pt}\checkmark & \checkmark \\ 
			Temperature balancing 				& $\times$	& $\times$	& $\times$    & $\times$		& $\times$ 	& $\times$	& $\times$     & \hspace{7pt}$\times$	& \hspace{4pt}$\times$     & \hspace{3pt}\checkmark & \hspace{3pt}\checkmark & \hspace{3pt}\checkmark & \checkmark \\
			State-of-health balancing 			& $\times$ 	& $\times$	& $\times$      & $\times$		& $\times$	& $\times$	& $\times$	& \hspace{7pt}$\times$     & \hspace{4pt}$\times$     & \hspace{3pt}$\times$ 	& \hspace{3pt}$\times$	& \hspace{3pt}$\times$	& \checkmark \\ 
			Optimal cell-level power control 		& $\times$ 	& $\times$	& $\times$	& $\times$	& $\times$	& $\times$	& $\times$     & \hspace{7pt}$\times$     & \hspace{4pt}$\times$     & \hspace{3pt}\checkmark & \hspace{3pt}\checkmark & \hspace{3pt}\checkmark & \checkmark \\ 
			Compatibility with cells of different types& $\times$ 	& $\times$	& $\times$	& $\times$	& $\times$	& $\times$	&\checkmark & \hspace{7pt}$\times$ & \hspace{4pt}\checkmark     & \hspace{3pt}$\times$ 	& \hspace{3pt}\checkmark	& \hspace{3pt}\checkmark & \checkmark \\
			Flexibility in series/parallel connection 	& L		& M 		&H 		&	H	& H 			& M 		& L 			& \hspace{7pt}L 			& \hspace{4pt}L 			& \hspace{3pt}$\times$ 	& \hspace{3pt}$\times$	& \hspace{3pt}$\times$	& H 			\\ 
			\hline\hline \\[-4mm]
			L: Low, M: Medium, H: High
		\end{tabular}
\end{table*}

Table~\ref{TABLE_1} summarizes the main BESS architectures in the literature for comparison. In this context, we propose the presented work to improve the state of the art.

\subsection{Statement of Contributions}

Despite  recent advances, the RBESS technology remains far from reaching a level of maturity in both design and control. The existing RBESS architectures use either only switches or only converters to enable good reconfigurability or good power regulation capability, but not both. There is also a lack of power control approaches sophisticated enough to maximize the operating performance of RBESS. To overcome these substantial shortcomings, our study presents the following specific contributions.

\begin{enumerate}[1)]
	\item We propose a new modular, reconfigurable power electronics architecture for RBESS. Differing from the existing ones, the proposed architecture, for the first time, integrates power switches with DC/DC converters for a combination of their respective merits, while using the fewest number of switches for each cell to our best knowledge. Among the various benefits that the proposed design brings are high reconfigurability---arbitrary bypass and parallel or series connection among the adjacent cells---and high regulatability in power supply from cell to system level to satisfy exogenous power demands, even under the occurrence of cell failures, and ensure equal use of the cells simultaneously.
	\item We develop a power management approach based on optimal control for the proposed RBESS to minimize the system-wide energy loss while supplying demanded power and equalizing the cells in SoC and temperature. Compared to the prior methods, our approach is different in two aspects. First, we adopt piecewise linear SoC/OCV approximation in the battery model convexification to enable control across low to high SoC regions. Second, previous methods may be subject to infeasibility when the cells' actual conditions make the pre-set operating constraints unsatisfiable. Here, we introduce slack variables to relax the optimization problem to ensure the feasibility and practical applicability of our approach. Finally, we present a reconfiguration method to modify the switching circuit topology after a faulty cell is isolated. 
	\item We develop an experimental prototype and conduct a series of experiments to validate and assess the performance of the proposed RBESS. The experiments involve various scenarios under non-uniform cell conditions, fault occurrence, and reconfiguration. The results demonstrate the effectiveness of the proposed architecture and power management approach.
\end{enumerate}

Based on the above contributions, our RBESS design presents significant advantages to distinguish itself from the literature, as shown in Table~\ref{TABLE_1}. A preliminary conference version of this work appeared in~\cite{9589321} to report the RBESS architecture design. Here, we introduce substantial expansions in optimal power management and experimental validation.

\subsection{Organization}

The remainder of the paper is organized as follows. Section II presents the power electronics architecture of the proposed RBESS and discusses its unique features. Section III develops the optimal power management approach, which covers the modeling, optimization problem formulation, and convexification. This section also presents a switching circuit reconfiguration method. Section IV provides extensive simulation results, and Section V proceeds to develop an experimental prototype to validate the proposed RBESS design and control approach. Finally, Section VI concludes the study.

\section{Architecture of the Proposed RBESS}

\begin{figure}[!t]\centering
	\includegraphics[width=\linewidth]{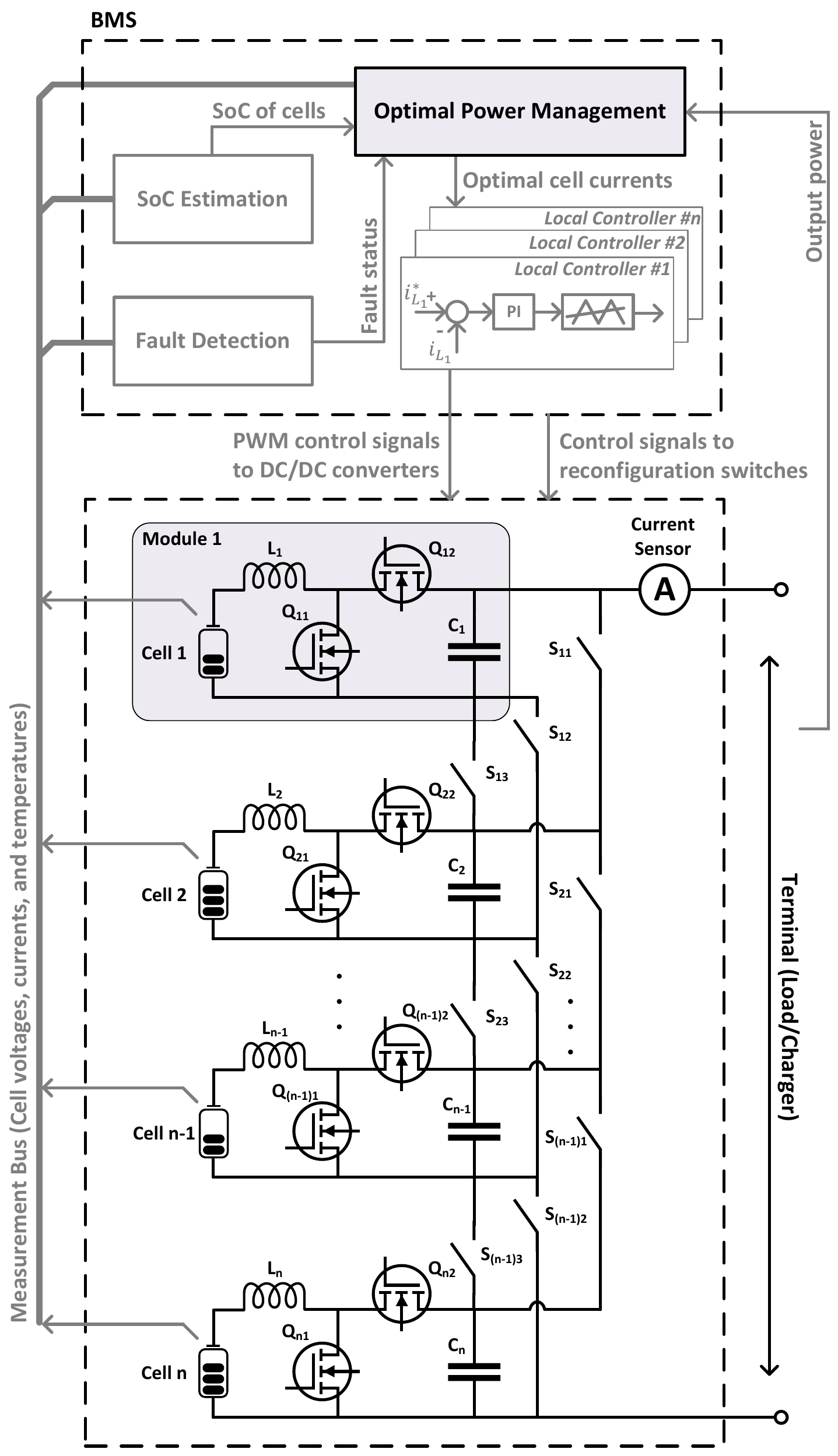}
	\caption{The proposed modular DC/DC converter-integrated RBESS.}\label{FIG_1}
\end{figure}

This section elaborates the proposed RBESS. Fig.~\ref{FIG_1} shows the power electronics architecture. As illustrated, every cell is connected with a DC/DC converter to make up a module. Here, we employ synchronous DC/DC converters, even though other types of converter topologies are also allowed. The converter, which comprises two power switches, an inductor, and an output capacitor, provides bi-directional power processing to control the charging and discharging of the cell. As such, the module has the capability of cell-level power control. The modules are connected via a switching circuit, which places three power switches between every two adjacent modules. With an appropriate reconfiguration of the switches, the switching circuit can bypass a module subject to faults and achieve arbitrary series or parallel connection among adjacent modules. We label the three switches connecting modules $i$ and $i+1$ as $S_{ij}$ for $j=1,2,3$. For switch $S_{ij}$, $S_{ij}=1$ when it is on, and $S_{ij}=0$ when it is off. To bypass and isolate cell $i$ for $1\leq i \leq n-1$ from the battery pack, we let $S_{i1}=1$ and $S_{i2}=S_{i3}=0$. To bypass cell $n$, we let $S_{(n-1)2}=1$ and $S_{(n-1)3}=S_{(n-1)1}=0$. Modules $i$ and $i+1$ are configured in series when $S_{i1}=S_{i2}=0$ and $S_{i3}=1$, and in parallel when $S_{i1}=S_{i2}=1$ and $S_{i3}=0$. 

The proposed design uses only $3(n-1)$ power switches for $n$ cells. To our knowledge, this is more economical than any other RBESS design in the literature to provide the same level of reconfigurability. The circuit simplicity further results in convenient operation and reconfiguration. Specifically, a cell requires only one switch to be on for the series and bypass configurations, and a parallel configuration needs only two switches to be on, as seen from above.

Fig.~\ref{FIG_1} illustrates how the proposed design dovetails with the battery management system (BMS). The BMS adopts a two-layer control strategy. At the higher level, the optimal power management generates optimal charging/discharging currents for the cells; at the lower level, PI-based local current mode controllers perform reference tracking for the cell currents, as is common in control of DC/DC converters \cite{4012131}.

Our proposed RBESS can provide high reconfigurability and control flexibility, which lead to distinct benefits for practical applications. A summary of the major ones is as follows.

\begin{enumerate}[1)]
	\item The proposed design allows to bypass and isolate any faulty module. Hence, the battery pack can continue to operate rather than shut down as a whole, despite safety threats and anomalies. Further, following the bypass of a module, the switching circuit can reconfigure to redirect the power flow and share the load equally among the remaining in-service cells to promote balanced use of them.
	\item In the proposed design, the embedded DC/DC converters take on responsibility for the external power electronic devices in the conventional designs to control the charging/discharging of the cells. The DC/DC converters would yield useful functions with their capability of power conversion and control. First, they can regulate their output voltage so that the RBESS can supply desired or reference voltage. The voltage supply can remain consistent before and after a fault-induced reconfiguration. Second, the converters apply individual current or voltage control to the cells, thus making it possible to customize and optimize the charging/discharging for each cell based on its present condition. One can translate this strength into cell balancing, e.g., by charging (resp., discharging)  the cells with high SoC less (resp., more) relative to the cells with low SoC. It is also viable to balance the cells in temperature and state-of-health.
	\item Even though beyond the scope of this paper, the proposed design can accommodate the heterogeneity of the cells. For example, one can leverage it to integrate heterogeneous cells from different manufacturers or even based on different electrochemistries to form a workable system. In a similar vein, the design can potentially enable hybrid energy storage consisting of battery cells, supercapacitors, and even solar cells.
\end{enumerate}

Note that the use of the embedded DC/DC converters can mitigate the use of exogenous power electronics devices, and that one can connect a string or pack of cells, rather than a single cell, to each DC/DC converter in practical adoptions. These factors would help make the overall implementation cost of the proposed RBESS manageable. Finally, the diverse new functions and safety improvements brought by the design are valuable especially for high-stakes, safety-critical applications including electric vehicles and aircraft.

\section{Modeling and Optimal Control of the Proposed RBESS}

This section investigates optimal power management for the proposed RBESS. We will begin with the electrical and thermal modeling for the modules. We then proceed to present a model-based optimal control problem for the RBESS and convexify it for computational tractability.  Here, the modeling and optimal control formulation are a refinement of the methodology in \cite{7544629,8768008}, and improvements are introduced to expand the operating SoC range of the RBESS and ensure the feasibility of the optimization problem. Finally, we present a switching circuit reconfiguration mechanism that dovetails with the power management method. 

\subsection{Electrical and Thermal Modeling}

\begin{figure}[!t]
    \centering
    \subfloat[\centering ]{{\includegraphics[width=7cm]{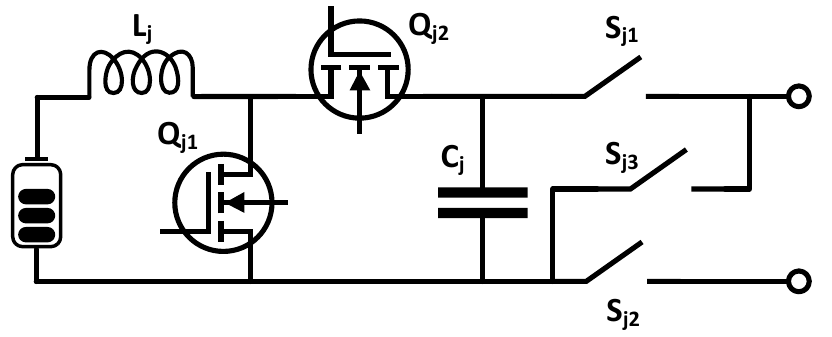} }}
    \qquad
    \subfloat[\centering ]{{\includegraphics[width=7cm]{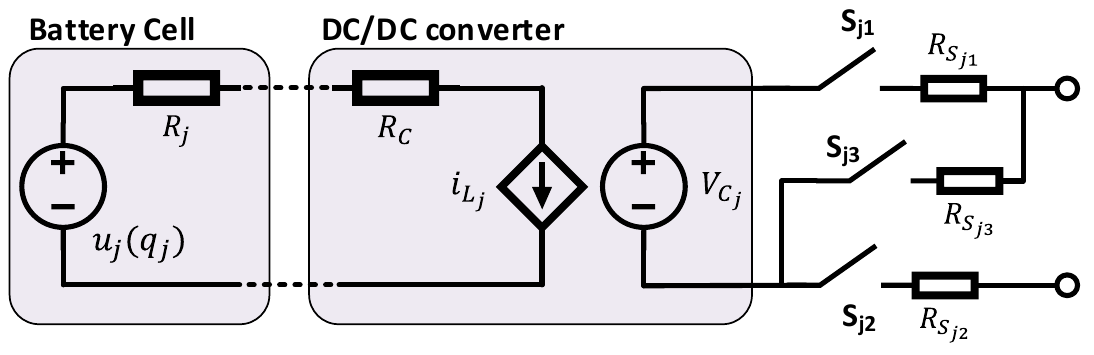} }}
	\qquad
    \subfloat[\centering ]{{\includegraphics[width=7cm]{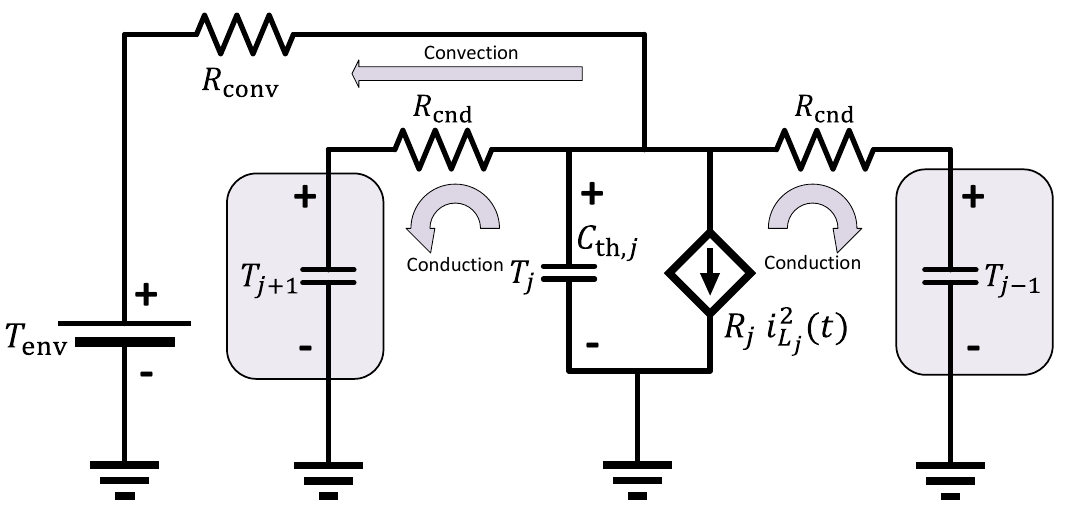} }}
    \caption{The proposed DC/DC converter-integrated cell module. (a) The circuit structure of the proposed module. (b) The electrical model of the proposed module. (c) The thermal model of the cell $j$.}
    \label{FIG_2}
\end{figure}

Consider the proposed RBESS consisting of $n$ modules. Each module includes a cell, a DC/DC converter, and three power switches connected in cascade, as shown in Fig.~\ref{FIG_2} (a). We denote the set of all the in-service modules as $\mathcal{J}$. The electrical model of module $j$ for $j \in \mathcal{J}$ is schematically shown in Fig.~\ref{FIG_2} (b). It includes two parts. The first part is the Rint model to describe the cell's electrical dynamics \cite{en4040582}, which comprises an OCV source $u_j$ in series with an internal resistor $R_j$. The model's governing equations are:
\begin{subequations}
\begin{align}
\dot{q}_j(t)&=-\frac{1}{\bar{Q}_j}i_{L_j}(t),\\
v_j(t)&=u_j(q_j(t))-R_ji_{L_j}(t),
\label{SoCDynamic}
\end{align}
\end{subequations}
where $v_j$, $i_{L_j}$, $u_j$, $\bar{Q}_j$, and $q_j$ are the terminal voltage of the cell, applied current in Ampere, OCV, capacity, and SoC, respectively. The internal power of the battery cell is given by
\begin{equation}
P_{b_j}=u_j(q_j(t))i_{L_j}(t).
\label{IntPower}
\end{equation}

The DC/DC converter is modeled as an ideal DC/DC transformer along with a series resistor $R_C$ to capture power losses. The reconfiguration switches are also modeled by the ideal switches with series resistors $R_{S_{ji}}$. For module $j$, we assume that we can collect the power losses on $R_{S_{ji}}$ for $i=1,2,3$ in a single resistor $R_{S_j}$. Thus, the module's output power $P_j$ can be calculated as
\begin{equation}
P_j=u_j(q_j(t))i_{L_j}(t)-(R_j+R_C+R_{S_{j}})i_{L_j}^2(t),
\label{ModulePower}
\end{equation}
where $R_ji_{L_j}^2(t)$, $R_Ci_{L_j}^2(t)$, and $R_{S_j}i_{L_j}^2(t)$ represent the internal power losses of the cell, the converter, and the reconfiguration switches, respectively.

We adopt a lumped thermal model in \cite{7544629} to describe the thermal dynamics of module $j$. The thermal model is depicted in Fig.~\ref{FIG_2} (c). The model captures the heat transfer due to the convection between module $j$ and the environment, $\dot{Q}_{\textrm{conv},j}$, and the conduction between module $j$ and its adjacent cells, $\dot{Q}_{\textrm{cnd},j}$. Meanwhile, the power loss caused by the internal resistor, $R_ji_{L_j}^2$, translates into heat generation, which becomes the main heating source. Combining all, the thermal model is governed by
\begin{subequations}
\label{TempDynamicAll}
\begin{align}
C_{\textrm{th},j}\dot{T}_j(t)&=R_ji_{L_j}^2(t)-\dot{Q}_{\textrm{cnd},j}-\dot{Q}_{\textrm{conv},j}, \label{TempDynamic}\\
\dot{Q}_{\textrm{conv},j}(t)&=(T_j(t)-T_{\textrm{env}})/R_{\textrm{conv}},\\
\dot{Q}_{\textrm{cnd},j}(t)&=(2T_j(t)-T_{j+1}(t)-T_{j-1}(t))/R_{\textrm{cnd}},
\end{align}
\end{subequations}
where $T_j$ and $T_{\textrm{env}}$ are the cell's and environmental temperatures, respectively. In addition, the term $C_{\textrm{th},j}$ represents the thermal capacitance of the cell; $R_{\textrm{cnd}}$ and $R_{\textrm{conv}}$ are the thermal resistances between neighboring cells and between cell $j$ and the environment, respectively. Here, $R_{\textrm{conv}}$ depends inversely on the external surface area of the cell $A_j$ and the convective heat transfer coefficient between the cell's surface and the environment $h$. For instance, one can consider a parallel forced air cooling approach for the proposed design \cite{PESARAN2002377}, which allows every cell to experience the same amount of cooling air.

The above electro-thermal model is concise but expressive and computationally efficient. Putting them together for all the modules, one can obtain a complete description of the dynamics of the RBESS, which allows us to perform optimal power management design subsequently. 

\subsection{Problem Formulation}

The aim of our RBESS power management is to distribute the power load among the cells so that the power losses can be minimized under some key safety, balancing and power demand satisfaction constraints. To begin with, we will formulate the optimization problem. 

The total power losses of the RBESS can be expressed by
\begin{equation}
J(t) = \sum_{j\in\mathcal{J}} (R_j+R_C+R_{S_j})i_{L_j}^2(t).
\label{Loss}
\end{equation}
We use the following objective function to encompass the total power losses over a horizon:
\begin{equation}
\int_{0}^{H} J(t)dt,
\end{equation}
where $H$ is the planning horizon length. For the sake of safety, we require each cell to operate within some favorable current, SoC, and temperature ranges:
\begin{subequations}
\begin{align}
i_{L_j}^\textrm{min}\leq i_{L_j}&\leq i_{L_j}^\textrm{max}, \label{SafetyConstraint}\\
q_j^\textrm{min}\leq q_j&\leq q_j^\textrm{max}, \label{SoCConstraint}\\
T_j&\leq T_j^{\textrm{max}},
\end{align}
\label{SafetyConstraints}
\end{subequations}
\hspace{-5pt}where $i_{L_j}^\textrm{min/max}$, $q_j^\textrm{min/max}$, and $T_j^{\textrm{max}}$ are the lower/upper safety bounds for the current, SoC, and temperature of cell $j$, respectively. It is important to note that $i_{L_j}^\textrm{min}$ can be set to be zero as the zero current means the bypass of the module. Further, we impose the following SoC and temperature balancing constraints to equalize the cells and make an even usage of them:
\begin{align}
|q_j(t)-q_{\textrm{avg}}(t)|& \leq \Delta q, \label{SoCBalancing} \\
|T_j(t)-T_{\textrm{avg}}(t)|& \leq \Delta T, \label{TempBalancing}
\end{align}
Here, $q_{\textrm{avg}}(t)$ and $T_{\textrm{avg}}(t)$ represent the average SoC and temperature of all the cells that belong to $\mathcal{J}$. They are calculated as
\begin{equation}
X_{\textrm{avg}}(t) = \frac{1}{\textrm{card}(\mathcal{J})}\sum_{j\in\mathcal{J}}^{}X_j(t),
\nonumber
\end{equation}
where $X=q$ and $T$, and $\textrm{card}(\mathcal{J})$ is the cardinality of $\mathcal{J}$. The SoC and temperature thresholds $\Delta q$ and $\Delta T$ determine the tolerated deviation of each cell's SoC and temperature from the average. Here, $\Delta q$ and $\Delta T$ are tunable parameters, and one can tune them to meet the SoC and temperature balancing requirements for a specifically considered application. While lower $\Delta q$ and $\Delta T$ values force a more balanced SoC and temperature distribution among the cells, higher values allow more deviation for the cells' SoC and temperature from the average. To make the RBESS meet the power demands, we present the following output power satisfaction constraint:

\begin{equation}
\sum_{j\in\mathcal{J}} P_{b_j}-(R_j+R_C+R_{S_j})i_{L_j}^2 = P_{\textrm{out}},
\label{DemandSatisfaction}
\end{equation}
where $P_{\textrm{out}}$ is the total power demanded of the RBESS.

Summing up the above, our power management approach is based on addressing the constrained nonlinear optimization problem as follows:
\begin{equation}
\begin{aligned}
\min_{i_{L_j}, j\in\mathcal{J}} \quad & \int_{0}^{H} J(t)dt,\\
\textrm{s.t.} \quad & \eqref{SoCDynamic},\eqref{TempDynamic}, \eqref{SafetyConstraints}-\eqref{DemandSatisfaction}.
\label{InitialOptim}
\end{aligned}
\end{equation}
This optimization problem pursues predictive minimization of the power losses while complying with the constraints that promote safety, SoC and temperature balancing, and power supply-demand match. Note that the optimization problem \eqref{InitialOptim} is non-convex due to the nonlinearity of the equality constraint \eqref{DemandSatisfaction}. Thus, the solution to this problem is neither trivial nor computationally cheap. To overcome the issue, we relax the problem slightly to formulate a convex optimization problem, as suggested in \cite{MURGOVSKI201292}. The convexification is described in detail as follows.

\subsection{Convex Problem Formulation}

\begin{figure}[!t]\centering
	\includegraphics[trim={3cm 0.5cm 3cm 1cm},clip,width=\linewidth]{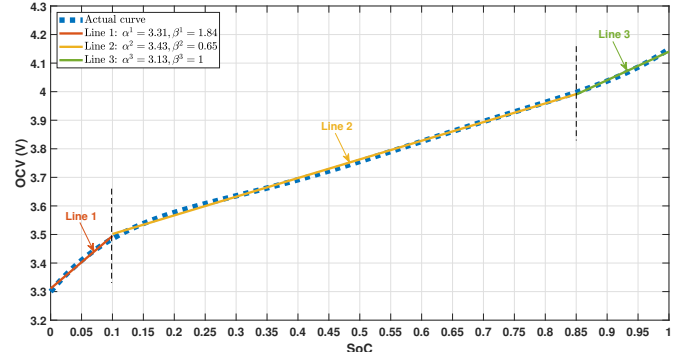}
	\caption{The SoC/OCV curve of the simulated cells and the multi-segment linearization.}\label{FIG_3}
\end{figure}

For the sake of convexification, we begin with linearizing the SoC/OCV curve. The existing studies, e.g., \cite{7544629,8768008} perform the linearization for only the medium SoC range, where the OCV is closely linear with SoC for lithium-ion batteries. However, this treatment excludes the use of the low and high SoC ranges. To address the issue, we introduce multi-segment linearization based on different SoC ranges to approximate the complete SoC/OCV curve:
\begin{equation}
u_j(q_j(t))=\alpha^i_j(q_j(t))+\beta^i_j(q_j(t))q_j(t),
\label{Linapprox}
\end{equation}
where $\alpha^i_j$ and $\beta^i_j$ are the intercept and slope coefficients of the $i$-th line segment for cell $j$. Fig.~\ref{FIG_3} illustrates an example of the linearization, where the SoC/OCV curve taken from a real cell is approximated by three line segments. Differing from the literature, the $\alpha^i_j$ and $\beta^i_j$ values are SoC-dependent, and the multi-segment linear approximation spans the SoC/OCV curve from 0 to 100\% SoC. To ease the notation, we will drop the superscript $i$ from $\alpha^i_j$ and $\beta^i_j$ in the sequel without causing confusion. Next, we present a convex model by introducing the notion of accumulated energy $E_j$ to take the place of SoC. The accumulated energy of a battery cell can be expressed as
\begin{equation}
E_j(t)=\frac{1}{2}C_ju_j^2(q_j(t))-E_j^0,
\label{Energy}
\end{equation}
where $C_j=\bar{Q}_j/\beta_j$ and $E_j^0=\frac{1}{2}C_ju_j^2(q_j(0))$ is the initial energy. Inserting \eqref{Linapprox} to \eqref{Energy} and using \eqref{SoCDynamic}, the dynamic equation of the cell's accumulated energy can be derived as
\begin{equation}
\dot{E}_j(t)=-P_{b_j}.
\label{EnergyDynamic}
\end{equation} 
In the above, we extract a desirably linear dynamic model to represent the evolution of $E_j(t)$ driven by $P_{b_j}$. Based on \eqref{EnergyDynamic}, we will reformulate the optimization problem to be one with respect to $P_{b_j}$, as will be seen later.
Proceeding forward, we consider module $j$'s power loss, $P_{l_j}(t)$, which can be expressed in terms of $P_{b_j}$ as
\begin{equation}
P_{l_j}(t)=\frac{(R_j+R_C+R_{S_j})C_jP_{b_j}^2(t)}{2(E_j(t)+E_j^0)}.
\label{PowerLossConvex}
\end{equation}
As our optimization goal is to minimize the total power loss, \eqref{PowerLossConvex} serves as an equality constraint. Since $P_{l_j}(t)$ is not a linear function of $P_{b_j}$, the resulting optimization problem would be non-convex due to the nonlinearity of \eqref{PowerLossConvex}. Thanks to the fact that the objective function minimizes the total loss of the battery pack, we can relax \eqref{PowerLossConvex} to comply with the convexity requirement
\begin{equation}
P_{l_j}(t) \geq \frac{(R_j+R_C+R_{S_j})C_jP_{b_j}^2(t)}{2(E_j(t)+E_j^0)},
\label{PowerLossConvexRelax}
\end{equation}
by which the optimization problem will practically reduce $P_{l_j}(t)$ to its lower bound. 
The safety constraints \eqref{SafetyConstraint}-\eqref{SoCConstraint} can also be reformulated in terms of the $P_{b_j}$ and $E_j$ as follows:
\begin{subequations}
\begin{align}
\sqrt{\frac{2}{C_j}(E_j+E_j^0)}i_{L_j}^\textrm{min} &\leq P_{b_j} \leq \sqrt{\frac{2}{C_j}(E_j+E_j^0)}i_{L_j}^\textrm{max}, \\
\frac{1}{2}C_ju_j^2(q_j^\textrm{min}(t)) &\leq E_j+E_j^0 \leq \frac{1}{2}C_ju_j^2(q_j^\textrm{max}(t)).
\end{align}
\end{subequations}
Similarly, the SoC balancing constraint \eqref{SoCBalancing} translates into the following one:
\begin{equation}
\Bigg|\frac{2}{C_j}E_j(t)-\frac{1}{\textrm{card}(\mathcal{J})}\sum_{i\in\mathcal{J}}^{}\frac{2}{C_i}E_i(t)\Bigg| \leq \Delta E_j,
\label{EBalancing}
\end{equation}
where $\Delta E_j=(\alpha_j+\beta_j\Delta q)^2-\alpha_j^2$. It is worth noting that the SoC balancing constraint, either \eqref{SoCBalancing} or \eqref{EBalancing}, may result in infeasibility for the optimization problem, when $\Delta q$ or $\Delta E$ fails to bound the cells' initial difference in SoC. The same issue applies to the temperature balancing constraint \eqref{TempBalancing}. Once happening, the infeasibility will cause the power optimization procedure to abort. While it is possible to make $\Delta E$ and $\Delta T$ large enough to forestall the issue, this will sacrifice the achievable performance in both power loss minimization and cell balancing. To guarantee the feasibility, we introduce slack variables to modify the constraints in \eqref{EBalancing} and \eqref{TempBalancing} as follows:
\begin{align}
\Bigg|\frac{2}{C_j}E_j(t)-\frac{1}{\textrm{card}(\mathcal{J})}\sum_{l\in\mathcal{J}}^{}\frac{2}{C_l}E_l(t)\Bigg|& \leq \Delta E_j+\xi^{(E)}_j,\\
|T_j(t)-T_{\textrm{avg}}(t)|& \leq \Delta T+\xi^{(T)}_j,
\end{align}
where $\xi^{(E)}_j, \xi^{(T)}_j \geq 0$ denote the SoC and temperature slack variables, respectively. The slack variables will be included into the objective function to penalize potential constraint violations. As such, if a cell's SoC or temperature is beyond the constraints, it will be driven close to the constraints by heavily penalizing the corresponding slack variables, without compromising the feasibility. Besides, the use of the slack variables will improve the power control flexibility. This will be discussed in the simulation study in Section IV.

Based on the above, we are now ready to make a convex relaxation of the problem in \eqref{InitialOptim}. Here, we also turn our focus to discrete-time optimization for the sake of computation, through applying the forward Euler method to \eqref{TempDynamicAll} and \eqref{EnergyDynamic} with the sampling time of $\Delta t$. We use the vector $z_j=[P_{b_j}, P_{l_j}, E_j, T_j, \xi_j^{(E)}, \xi_j^{(T)}]^\top, j\in \cal{J}$ to collect all the optimization variables and propose the following convex optimization problem for the RBESS power management:
\begin{equation}
\begin{aligned}
&\min_{z_j, j\in\mathcal{J}} \quad \sum_{k=0}^{H} \sum_{j\in\mathcal{J}} P_{l_j}[k] +\lambda^{(E)} \xi^{(E)}_j[k] + \lambda^{(T)} \xi^{(T)}_j[k],\\
&\textrm{Safety constraints:} \quad \\
&\ \sqrt{\frac{2}{C_j}(E_j[k]+E_j^0)}i_{L_j}^\textrm{min} \leq P_{b_j}[k] \leq \sqrt{\frac{2}{C_j}(E_j[k]+E_j^0)}i_{L_j}^\textrm{max},\\
&T_j\leq T_j^{\textrm{max}}\\
&\ \frac{1}{2}C_ju_j^2(q_j^\textrm{min}[k]) \leq E_j[k]+E_j^0 \leq \frac{1}{2}C_ju_j^2(q_j^\textrm{max}[k]),\\
&\textrm{Balancing constraints:} \quad \\
&\ \Bigg|\frac{2}{C_j}E_j[k]-\frac{1}{\textrm{card}(\mathcal{J})}\sum_{l\in\mathcal{J}}^{}\frac{2}{C_l}E_l[k]\Bigg| \leq \Delta E_j+\xi^{(E)}_j[k],\\
&\ |T_j[k]-T_{\textrm{avg}}[k]| \leq \Delta T+\xi^{(T)}_j[k],\\
&\textrm{Power loss constraint:} \quad \\
&\ P_{l_j}[k] \geq \frac{(R_j+R_C+R_{S_j})C_jP_{b_j}^2[k]}{2(E_j[k]+E_j^0)},\\
&\textrm{Energy dynamics:} \quad \\
&\ E_j[k+1]-E_j[k]=-P_{b_j}[k]\Delta t, \\
&\textrm{Thermal dynamics:} \quad \\
&\ T_j[k+1]=T_j[k] + \frac{\Delta t}{C_{\textrm{th},j}}\Big[P_{l_j}[k]-(T_j[k]-T_{\textrm{env}})/R_{\textrm{conv}}\\
& \quad -(2T_j[k]-T_{j+1}[k]-T_{j-1}[k])/R_{\textrm{cnd}}\Big],\\
&\textrm{Power supply-demand balance:} \quad \\
&\ \sum_{j\in\mathcal{J}} P_{b_j}[k]-P_{l_j}[k]=P_{\textrm{out}}[k],
\label{ConvexOptim}
\end{aligned}
\end{equation}
where $\lambda^{(E)}$ and $\lambda^{(T)}$ are the respective penalty weights for $\xi^{(E)}$ and $\xi^{(T)}$. The above problem is verifiably convex as a result of the convex cost function and constraints. The convexity makes it advantageous in practice as robust algorithms are available to find out its global optimum with efficient computation. The problem setup is similar to~\cite{7544629,8768008}. However, we crucially introduce the slack variables here to make the problem always feasible. This improvement eliminates the risk of no solution to satisfy the hard constraints, further enhancing the  practical aspects of power management.

The   problem is designed to be implemented in a receding-horizon manner. This will bring three benefits. First, predictive optimization over a limited time horizon rather than the whole mission duration will make the computation more manageable. Second, the receding-horizon power control can better respond to changes that occur to the RBESS in operation, e.g., fault-triggered cell bypass and switching circuit reconfiguration. Finally, the SoC change in each receding horizon is slight, so the optimization only needs to consider a single SoC/OCV linear segment and hence runs more efficiently.

\subsection{Reconfiguration}

\begin{algorithm}[!t]
 \caption{Power management of the proposed RBESS}
 \label{algorithm}
\begin{algorithmic}[1]  
  \FOR {Run-time}
	  \IF {a fault occurs to cell $i$}
	  \STATE Bypass cell $i$
	  \STATE Determine the set of the in-service cells $\mathcal{J}$
	  \STATE Calculate $n_s$ and $n_p$ from \eqref{Reconfiguration}
	  \STATE Reconfigure the switch circuit $S_{i1:3}, i=1,2,...n$
	  \ENDIF
  \STATE Run the optimal power management strategy \eqref{ConvexOptim}
  \RETURN $P_{b_j}$ 
	  \IF {$P_{b_j}==0$ for any $j$}
	  \STATE Bypass cell $j$
	  \ENDIF
  \ENDFOR
 \end{algorithmic}
 \end{algorithm}

The proposed RBESS allows dynamic switching of the power switches to bypass faulty cells, ensuring continuous system operation. Following the bypass, an important question is how to reconfigure the connection topology among the cells. However, it is not easy to identify a complete answer, as the large discrete reconfiguration decision space due to the use of switches would defy an exhaustive search for an optimal topology. In addition, inappropriate reconfiguration may produce poor topologies to cause short circuits or other issues. Note that the power management approach in \eqref{ConvexOptim} determines the optimal charging/discharging power of the cells individually and is not affected by any arbitrary series or parallel connection among them. This makes its run and the reconfiguration procedure separable but contiguous.

Here, we leverage an efficient heuristic to address the question and outline it as below. Suppose that all the remaining cells are approximately uniform in SoC and temperature at the time of the reconfiguration, since the power management based on \eqref{ConvexOptim} has driven cell balance. The reconfiguration then should yield a topology that facilitates a balanced use of the cells and makes every cell take an even power load. A straightforward topology design to fulfill this need is one based on $n_s$ serially connected modules with each module consisting of $n_p$ cells in parallel connection. We denote this topology as $n_p{\mathrm P}n_s{\mathrm S}$. We can determine $n_s$ and $n_p$ by
\begin{equation}
n_s=\frac{V_t^*}{V_C^\textrm{max}}, \qquad n_p=\frac{I_{\textrm{out}}}{i_C^\textrm{max}},
\label{Reconfiguration}
\end{equation}
where $V_t^*$ is the desired terminal voltage; $V_C^\textrm{max}$ and $i_C^\textrm{max}$ are the maximum output voltage and current stresses of the DC/DC converters, and $I_{\textrm{out}}=P_{\textrm{out}}/V_t^*$ is the output charging/discharging current of the battery pack. Subsequently, the RBESS can follow the series/parallel switching analysis in Section II to reconfigure the switch circuit.

This heuristic-based reconfiguration mechanism is computationally fast, fail-safe, and easy to implement. Further, it promotes system-wide cell balance and fits together with the power management in \eqref{ConvexOptim}. This leads us to the overall RBESS management approach as shown in Algorithm~\ref{algorithm}.

\section{Simulation Results}

\begin{table}[!t]
	\renewcommand{\arraystretch}{1.3}
	\caption{Specifications of the Proposed RBESS}
	\centering
	\label{TABLE_2}
	\resizebox{\columnwidth}{!}{
		\begin{tabular}{l l l}
			\hline\hline \\[-3mm]
			\multicolumn{1}{c}{Symbol} & \multicolumn{1}{c}{Parameter} & \multicolumn{1}{c}{Value [Unit]}  \\[1.6ex] \hline
			$ n $ & Number of battery cells & 15 \\
			$ v $  & Cell nominal voltage & 3.6     [V] \\
			$ \bar{Q} $ & Cell nominal capacity & 2.5     [A.h] \\ 
			$ R $ & Cell internal resistance & 31.3     [m$\Omega$] \\
			$ [q^{\textrm{min}},q^{\textrm{max}}] $ & Cell SoC limits  & [0.05,0.95] \\ 
			$ [i^{\textrm{min}},i^{\textrm{max}}] $ & Cell current limits & [-10,10]     [A] \\ 
			$ v_{\textrm{cut-off}} $ & Cell cut-off voltage & 3.3    [V] \\
			$ C_{\textrm{th}} $ & Thermal capacitance & 40.23     [J/K] \\ 
			$ R_{\textrm{conv}} $ & Convection thermal resistance & 41.05     [K/W] \\ 
			$ R_{\textrm{cnd}} $ & Conductance thermal resistance & 26.6     [K/W] \\ 
			$ T_{\textrm{env}} $ & Environment temperature & 298     [K] \\ 
			$ \Delta q $ & SoC balancing threshold & 1\% \\
			$ \Delta T $ & Temperature balancing threshold & 0.5    [K] \\
			$ \Delta t $ & Sampling time & 1     [s]\\
			\hline\hline
		\end{tabular}
	}
\end{table}

This section presents simulation results to evaluate the proposed RBESS design and power management approach. Table~\ref{TABLE_2} summarizes the specifications of the RBESS under simulation. The battery cells are assumed to be Samsung INR18650-25R, and we have identified their parameters (see Table~\ref{TABLE_2}) and SoC/OCV relationship (see Fig.~\ref{FIG_3}) from experiments using the approach in \cite{TIAN2020101282}. We approximate the SoC/OCV curve using a piecewise linear function with three segments that together span from zero to 100\% SoC. The power load profile for $P_{\textrm{out}}$ is obtained by repeating the scaled Urban Dynamometer Driving Schedule (UDDS). We use the CVX package \cite{cvx} to configure and solve the convex optimization problem in \eqref{ConvexOptim} to compute $P_b$. The optimization runs over a receding horizon of 20 seconds, i.e., $H=20$.

The initial SoC of the cells is drawn from a normal distribution with mean of 90\% and variance of 3\%. Similarly, the initial temperature of the cells follows a normal distribution with mean of 308 K and variance of 3 K. In order to investigate whether the power management can handle the cells' heterogeneity, a white Gaussian noise with variance of 4 m$\Omega$ is added to the internal resistance value of each cell. Furthermore, it is assumed that cells 4, 8, and 14 are bypassed and isolated from the battery pack at the 2,000th, 4,000th, and 6,000th seconds, respectively.

\begin{figure*}[t]
	    \centering
    \subfloat[\centering ]{{\includegraphics[trim={3cm 0 3cm 1cm},clip,width=8.5cm]{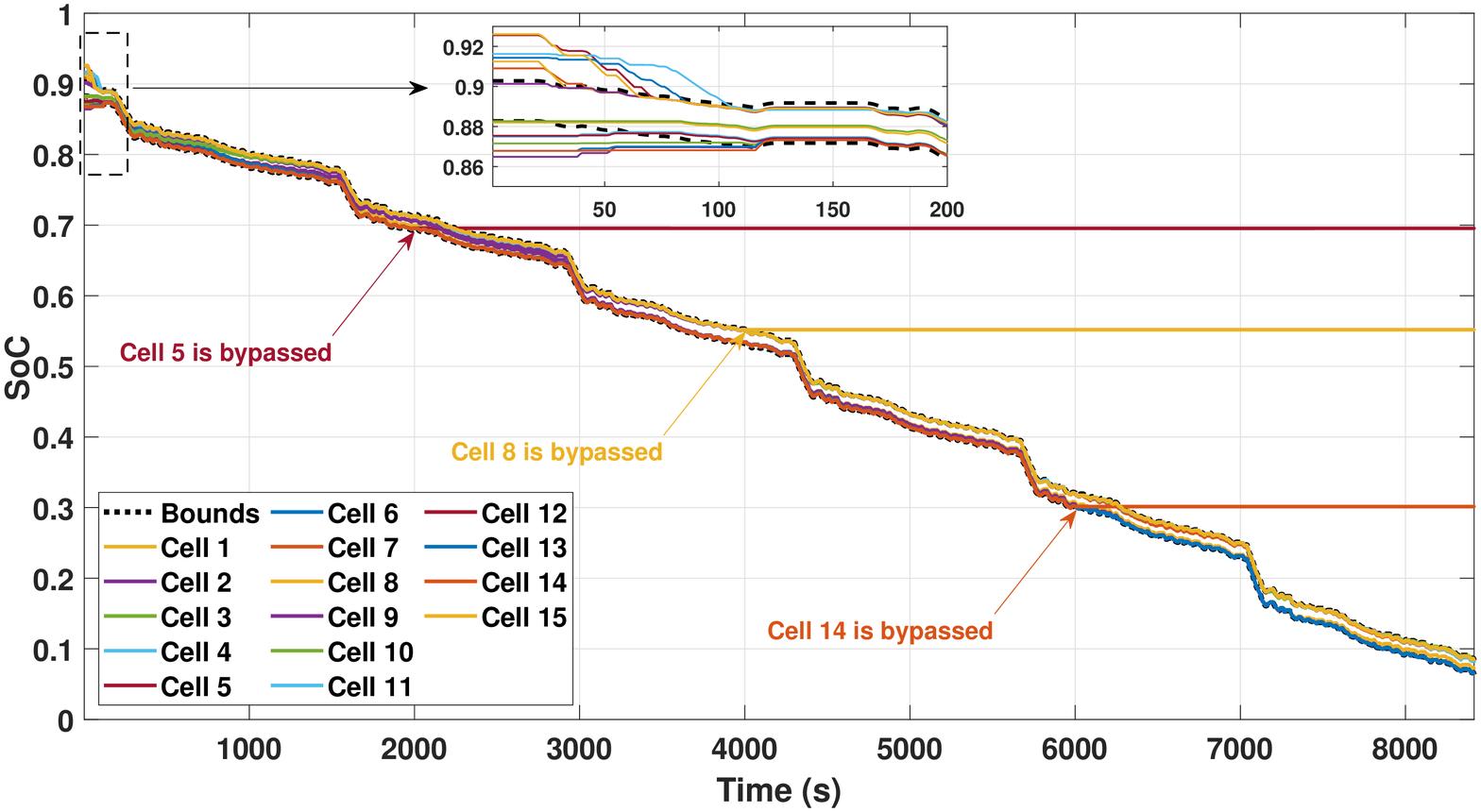} }}
	\,
    \subfloat[\centering ]{{\includegraphics[trim={2.5cm 0 3cm 1cm},clip,width=8.5cm]{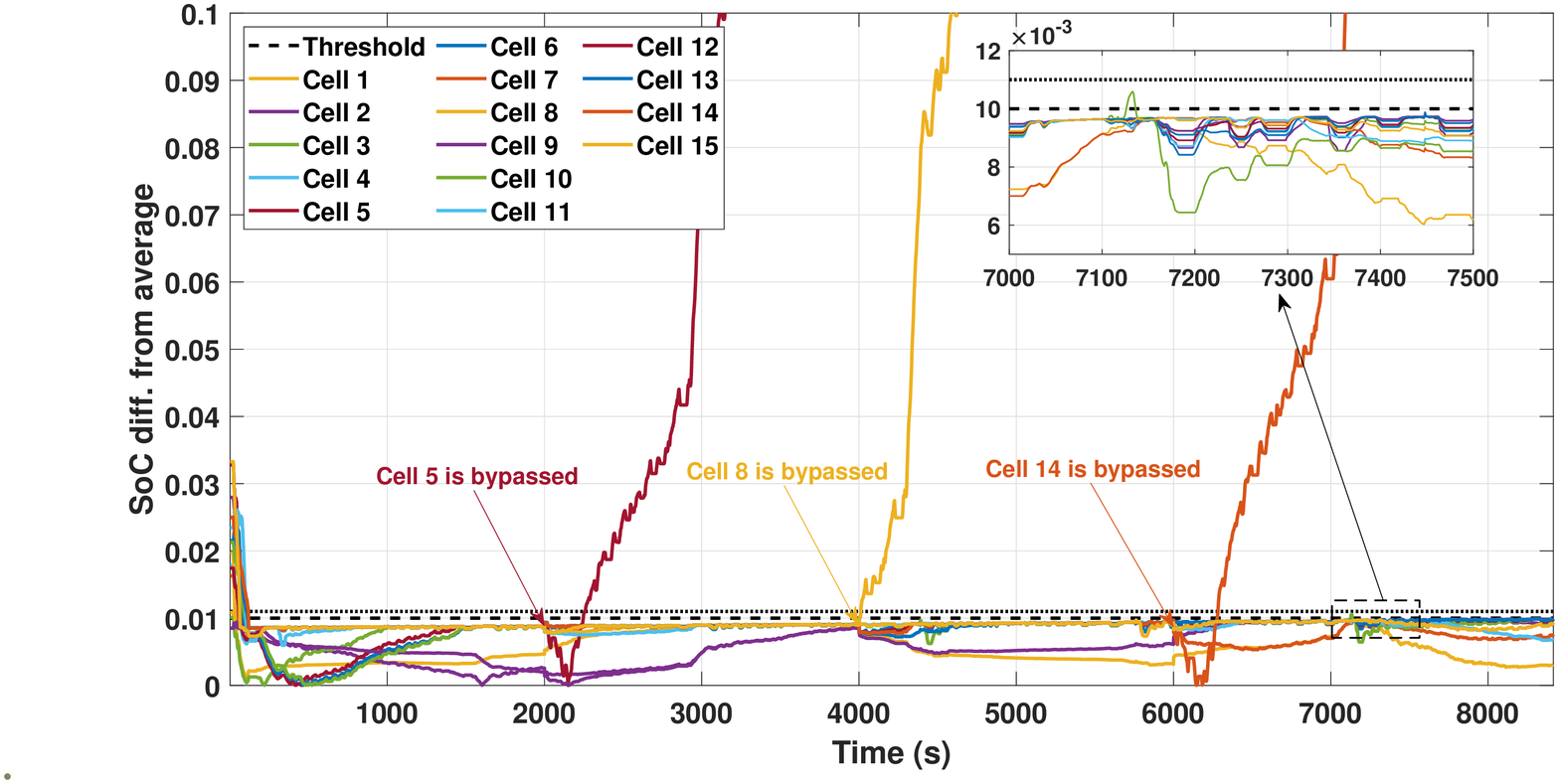} }}
    \,
    \subfloat[\centering ]{{\includegraphics[trim={3cm 0 3cm 1cm},clip,width=8.5cm]{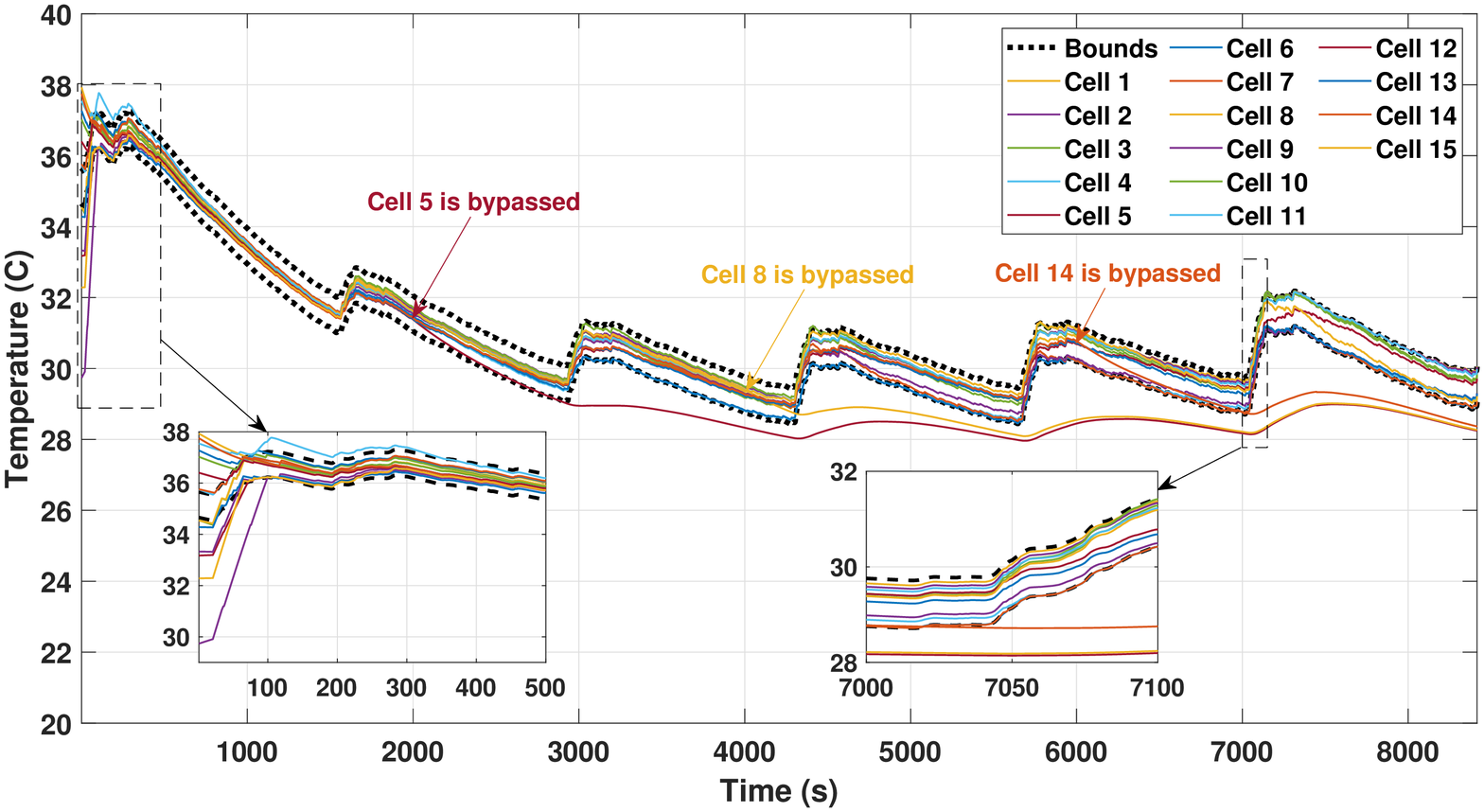} }}
    \,
    \subfloat[\centering ]{{\includegraphics[trim={2.5cm 0 3cm 1cm},clip,width=8.5cm]{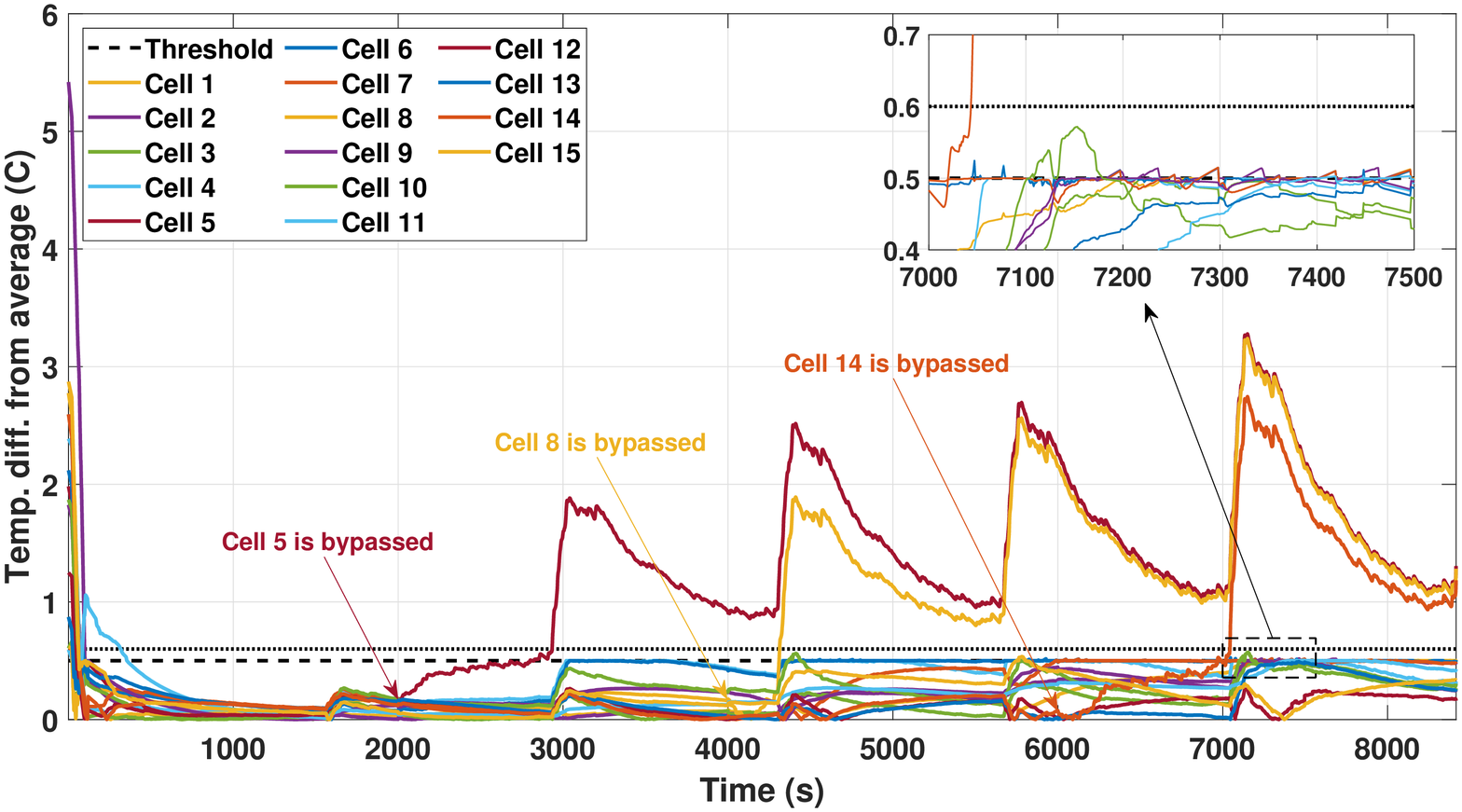} }}
    \caption{Simulation results of the SoC and temperature balancing. (a) The SoC of the cells. (b) The SoC difference of the cells from the average. (c) The temperature of the cells. (d) The temperature difference of the cells from the average.}
    \label{FIG_4}
\end{figure*}

\color{black}Fig.~\ref{FIG_4} depicts the SoC and temperature balancing performance of the proposed power management approach. The tolerated SoC and temperature deviation bounds, $\Delta q$ and $\Delta T$, are 1\% and 0.5 K, respectively. According to Fig.~\ref{FIG_4} (a), the cells are different in their initial SoC. Among them, cell 2 has the lowest initial SoC of 86.48\%, and cell 15 has the highest SoC of 92.61\%. The difference is beyond the desired error bounds. However, the power management approach successfully drives the SoC of the cells to reach within the bounds after 200 seconds and continues to regulate the charging/discharging power of the cells to ensure  SoC balance in the battery pack. Both cells 2 and 15 end up with the same SoC of 8.2\% when the simulation is finished. It is important to note the key of incorporating the slack variables in guaranteeing the feasibility of the power optimization. Fig.~\ref{FIG_4} (b) illustrates the deviation of the cells' SoC from the average. The tolerance bound is set to be 1\%. Some of the cells initially are beyond this bound---for example, cell 15 deviates from the average SoC by 3.5\%. In this case, the optimization problem would have been infeasible, but this issue is avoided as the slack variable $\xi^{(E)}$ permits slight violation of the SoC balancing constraints with a negligible compromise to physical safety of the cells. Meanwhile, the penalization of $\xi^{(E)}$ as in \eqref{ConvexOptim} in the cost function forces the cells to remain within the tolerated error bound once after they enter the bound, keeping the SoC balanced. The SoC of the bypassed cells remains unchanged after isolation as the cells are no longer used.

Fig.~\ref{FIG_4} (c) shows the evolution of the cells' temperature. Similar to the SoC initialization, the initial temperatures of the cells stretch beyond the desired bounds where cells 1 and 2 have the highest and lowest temperature of 37.92$\tccentigrade$ and 29.73$\tccentigrade$, respectively. The power management approach effectively controls the cell temperatures to reach a balanced temperature after 500 seconds. Note that a cell's temperature is still affected by the temperature of its adjacent cells and the environment after it is bypassed. As shown in Fig.~\ref{FIG_4} (c), when the average temperature of the battery pack increases, the temperature of the bypassed cells also rises due to the conductive heat transfer among adjacent cells. Here, even though the cells' initial temperature difference exceeds the bound of 0.5$\tccentigrade$ (see Fig.~\ref{FIG_4} (d)), the optimization for temperature balancing maintains feasibility as a result of introducing the slack variable $\xi^{(T)}$.

\begin{figure}[!t]\centering
	\includegraphics[trim={2.6cm 0.5 0.5cm 1cm},clip,width=\linewidth]{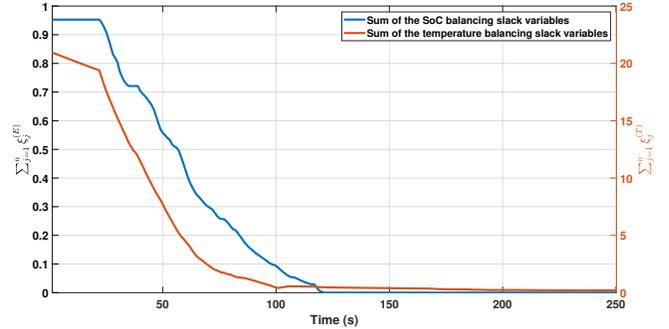}
	\caption{The change of the slack variables.}\label{FIG_5}
\end{figure}

To further investigate the role of the slack variables in the formulated optimization problem, Fig.~\ref{FIG_5} depicts their evolution through time. When the cells' SoC or temperature lies outside the balancing constraints, the slack variables will take nonzero values to relax the balancing constraints gently, thus turning the nominally infeasible optimization problem into a feasible one. The slack variables will decrease and approach zero as the cells are increasingly balanced in SoC and temperature. When they are zero, the SoC and temperature balancing constraints are fully satisfied. Penalizing the slack variables restricts over-relaxation of the constraints and tightens the bounds as the SoC and temperature get closer to or into the constraints. The penalization weights associated with the slack variables are subject to tuning so as to achieve the performance desired by a user. In general, heavier penalization will lead to less constraint relaxation and more time to achieve balancing. 

\begin{figure}[!t]\centering
	\includegraphics[trim={2.8cm 0cm 3cm 1cm},clip,width=\linewidth]{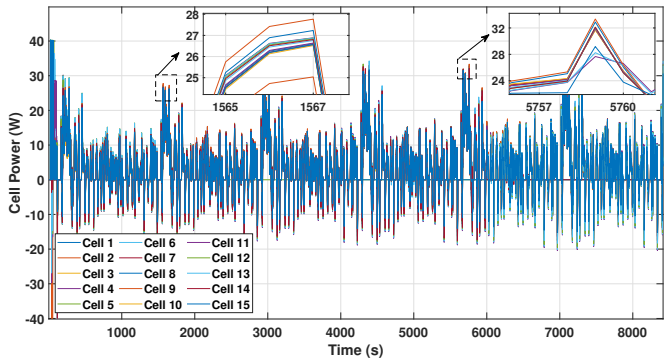}
	\caption{The output power profiles of the cells.}\label{FIG_6}
\end{figure}

The output power profiles of the cells are shown in Fig.~\ref{FIG_6}. We can see that the power of the individual cells is regulated to vary from one to another. This is because the cells have different conditions in SoC, temperature and internal resistances and must collectively minimize the overall power losses while complying with safety and balancing constraints. The cells can also adjust their own output on the bypass of a faulty cell. The peak power of battery cells are around 28 W for $1,560<t<1,570$ s before any cells are bypassed from the pack. However, when three cells are bypassed from the pack, the peak power of the remaining cells is increased to around 33 W for $5,750<t<5,760$ s to compensate for the bypassed cells and to ensure a continuous power supply to the load.

\begin{figure}[!t]\centering
	\includegraphics[trim={2.8cm 0cm 2.8cm 1cm},clip,width=\linewidth]{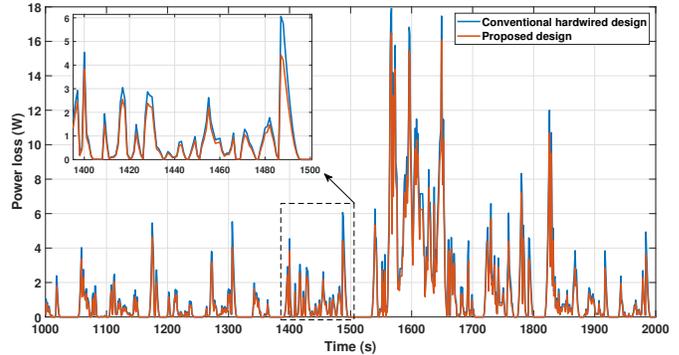}
	\caption{The power loss comparison.}\label{FIG_7}
\end{figure}

It is of our interest to investigate whether the proposed RBESS is more capable of reducing the total power losses than conventional hardwired battery systems. Fig.~\ref{FIG_7} shows a comparison of the resultant power losses, which focuses on $1,000<t<2,000$ s for the purpose of visual illustration. The hardwired pack is found to constantly suffer more power dissipation, because the it neglects that the cells have different internal resistances (associated with the higher power losses, the pack also faces higher operating temperatures as well as significant SoC and temperature imbalance). By contrast, the proposed RBESS is able to optimally allocate the charging/discharging power among the cells to gain more power efficiency.

\begin{figure}[!t]\centering
	\includegraphics[trim={3cm 0cm 3cm 1cm},clip,width=\linewidth]{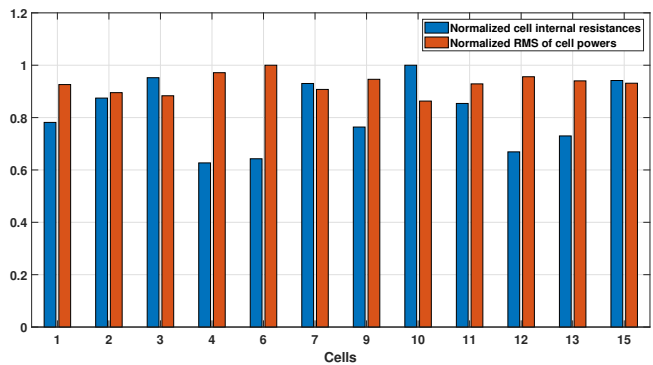}
     \captionsetup{belowskip=-15pt}
	\caption{The normalized internal resistance and RMS of the output power of the cells.}\label{FIG_8}
\end{figure}

We further assess whether the proposed power management approach can distribute power among the cells relative to their state-of-health (SoH), which is important to reduce the cell aging and degradation. To this end, we consider the root-mean-square (RMS) of the output power of the cells, and use the internal resistance as the SoH indicator---overall, the higher the internal resistance, the more degraded the cell is. Fig.~\ref{FIG_8} illustrates the normalized RMS of the output power of the battery cells in comparison to their internal resistance values. We observe that the cells with lower resistances are allocated more power load overall, see the groups of cells 1-4, cells 6-7, and cells 9-10. While the pattern is obvious, the power distribution also depends on each cell's SoC and temperature and thus shows certain perturbations. We can argue that the power management approach contributes to a balanced use of the battery cells in terms of SoH. 

\section{Experimental Results}

\begin{figure}[!t]
    \centering
    \subfloat[\centering ]{{\includegraphics[width=8.5cm]{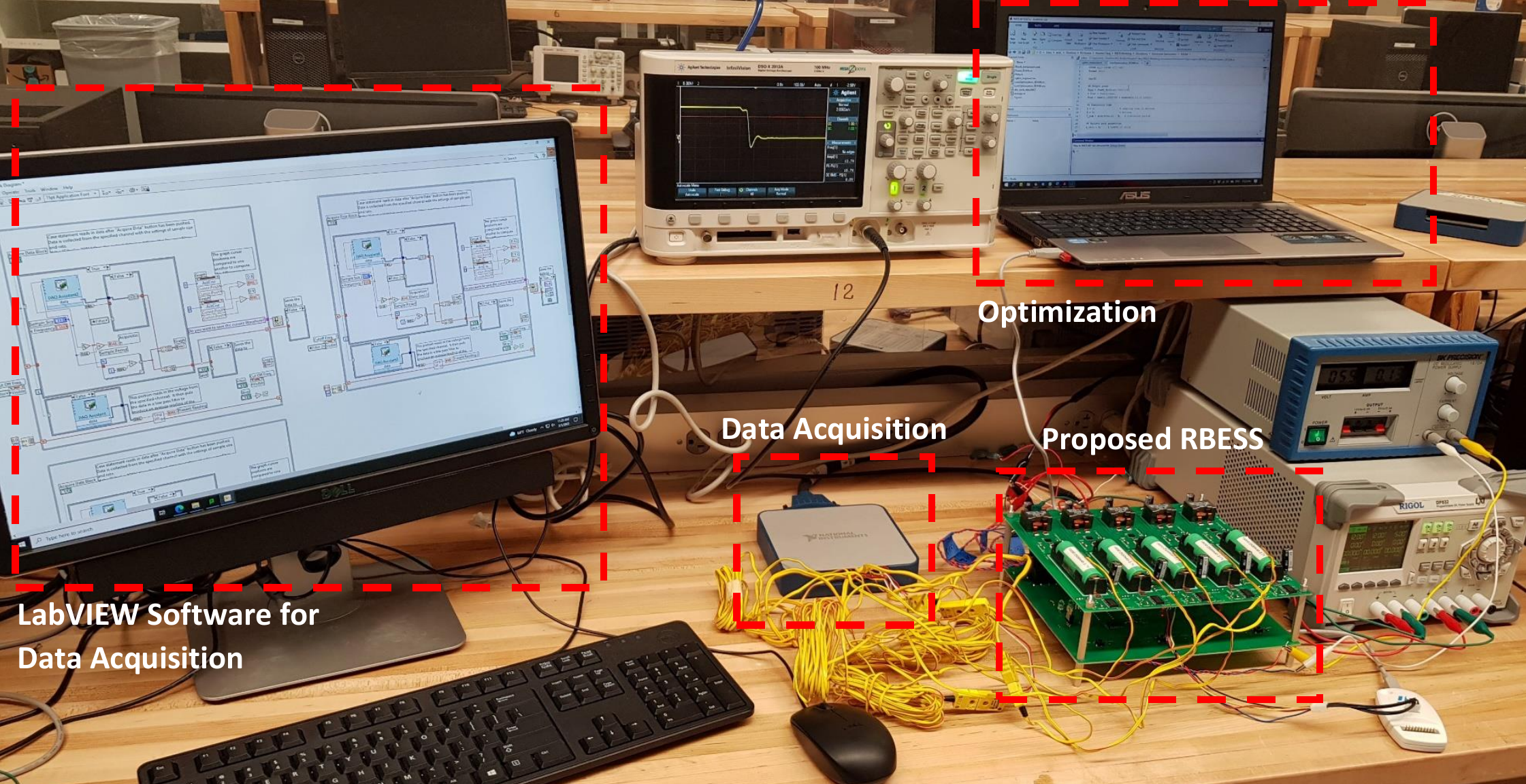} }}
    \qquad
    \subfloat[\centering ]{{\includegraphics[width=3.7cm]{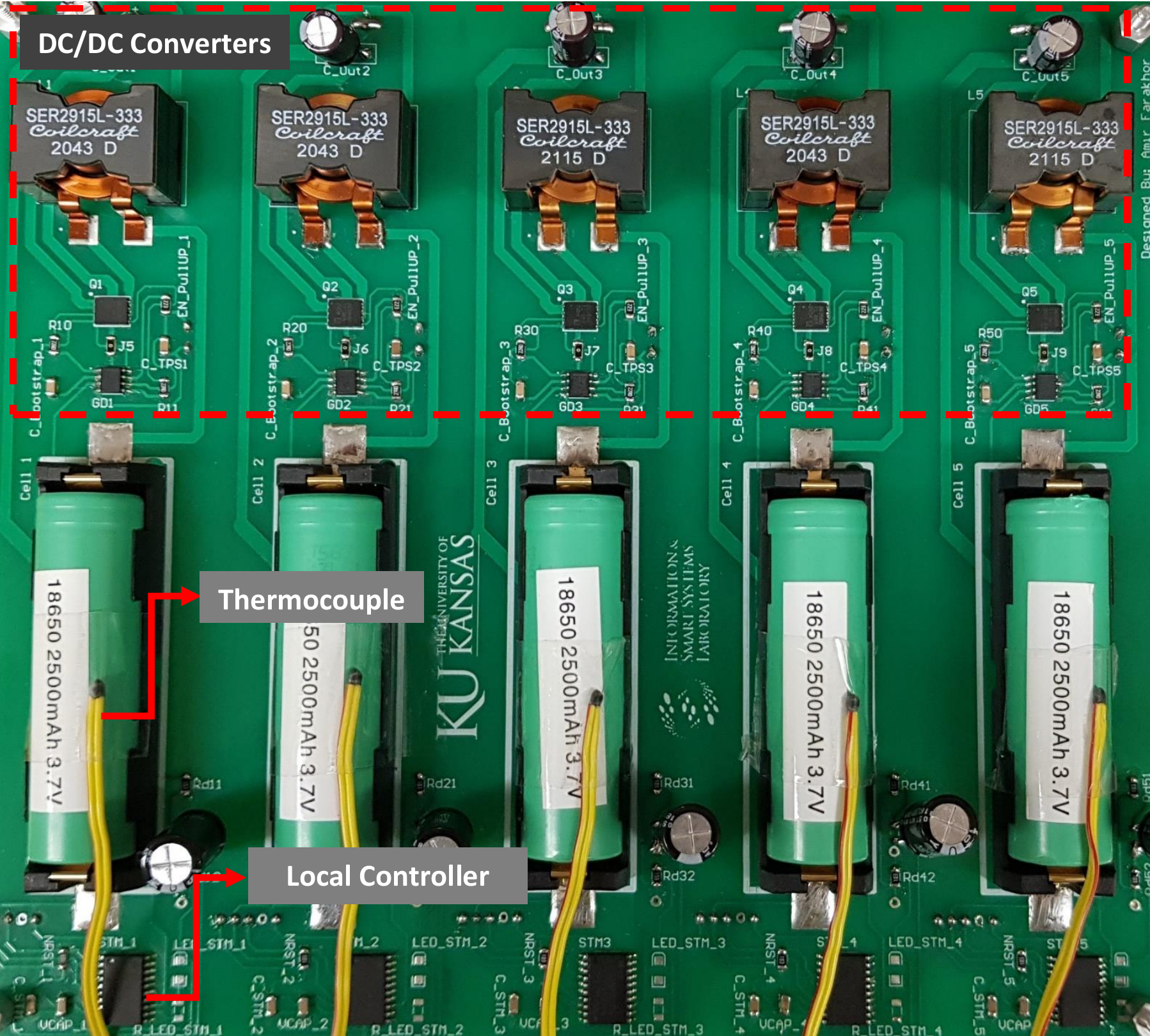} }}
\qquad
    \subfloat[\centering ]{{\includegraphics[width=3.9cm]{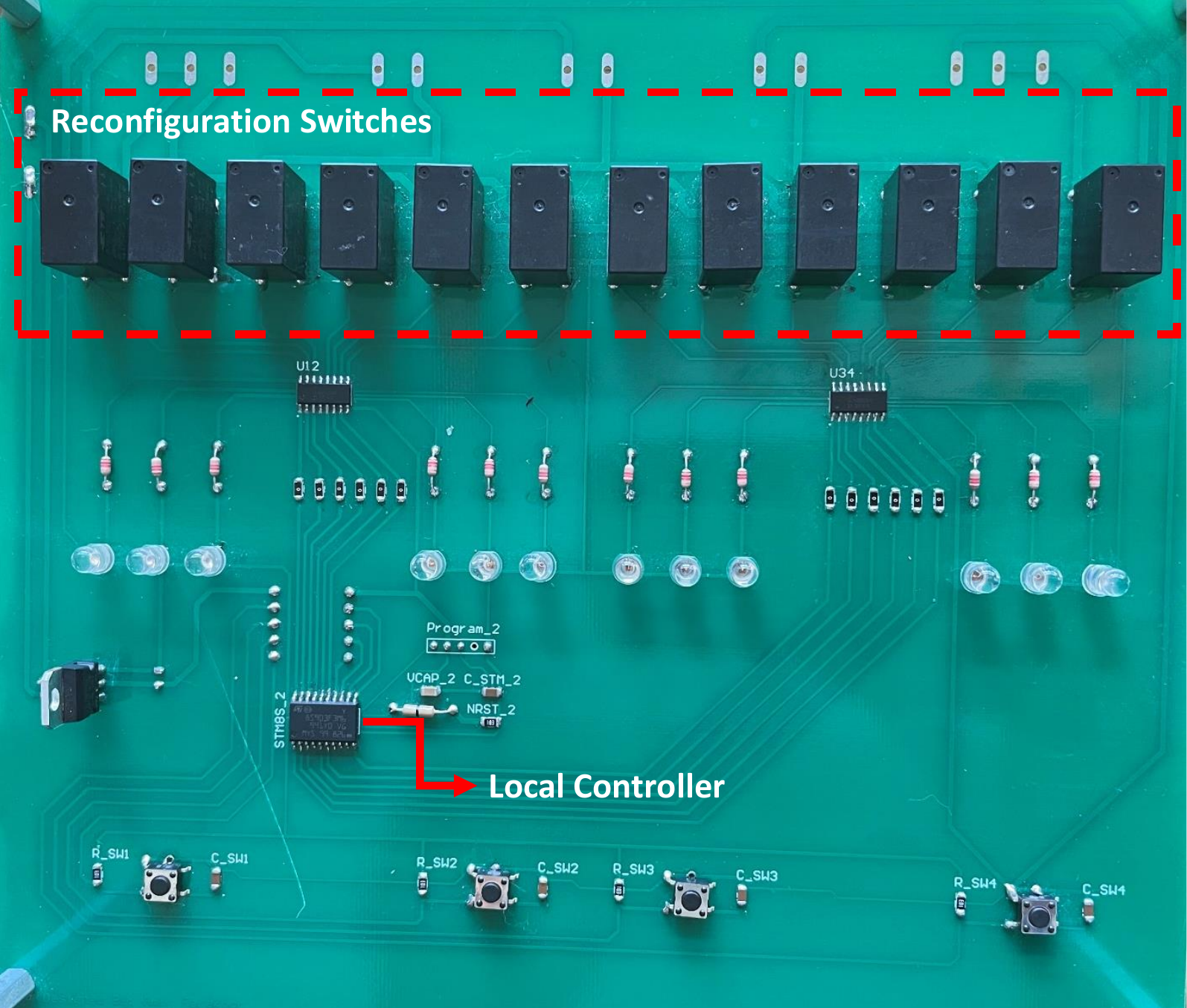} }}
    \caption{Lab-scale prototype of the proposed RBESS. (a) The experimental setup. (b) Top circuit view. (c) Bottom circuit view}
    \label{FIG_9}
\end{figure}

We develop a lab-scale prototype of the proposed RBESS for experimental validation. Fig.~\ref{FIG_9} (a) shows the experimental setup, and Figs.~\ref{FIG_9} (b)-(c) illustrate the circuit boards of the RBESS prototype based on the design in Fig.~\ref{FIG_1}. The RBESS is a pack of five cells integrated with five converters (Fig.~\ref{FIG_9} (b)) and 12 relay switches for reconfigurable connection (Fig.~\ref{FIG_9} (c)). Table~\ref{TABLE_3} lays out the specifications of the key components of the prototype. Type K thermocouples are attached on the surface of each cell to measure their temperature. A National Instruments PCIe-6321 DAQ board with LabVIEW is used to collect the cells' voltages, temperatures, and output power data. Using the CVX package, we then solve the optimal power management problem using MATLAB every minute (i.e., $\Delta t=60$ s). The optimal power values of the cells are then fed to local controllers using DSP TMS320F28335. The local controllers based on STM8S003F3P6 microcontrollers, generate 250 kHz PWM signals to DC/DC converters. The prototype is connected to a 20 $\Omega$ resistance load with a total output discharge power of 50 W.

\begin{table}[!b]
	\renewcommand{\arraystretch}{1.3}
	\caption{List of Key Components}
	\centering
	\label{TABLE_3}
	\centering
	{
		\begin{tabular}{l l}
			\hline\hline \\[-3mm]
			\multicolumn{1}{c}{Device} & \multicolumn{1}{c}{Model (Value)}\\[1.6ex] \hline
			MOSFET & CSD86356Q5D\\
			Relay switch & TE OJT-SS-105HM\\
			Gate driver & TPS28225\\
			Inductor & SER2915H-333KL (33 $\mu$H) \qquad \qquad \\
			Capacitor & (10 $\mu$F)\\
			Local controller \qquad \qquad \qquad & STM8S003F3P6\\
			Main controller & TMS320f28335\\
			Battery cell & Samsung INR18650-25R\\
			\hline\hline
		\end{tabular}
	}
\end{table}

\begin{figure*}[t]
	    \centering
    \subfloat[\centering ]{{\includegraphics[trim={2.8cm 0 2.8cm 1cm},clip,width=8.5cm]{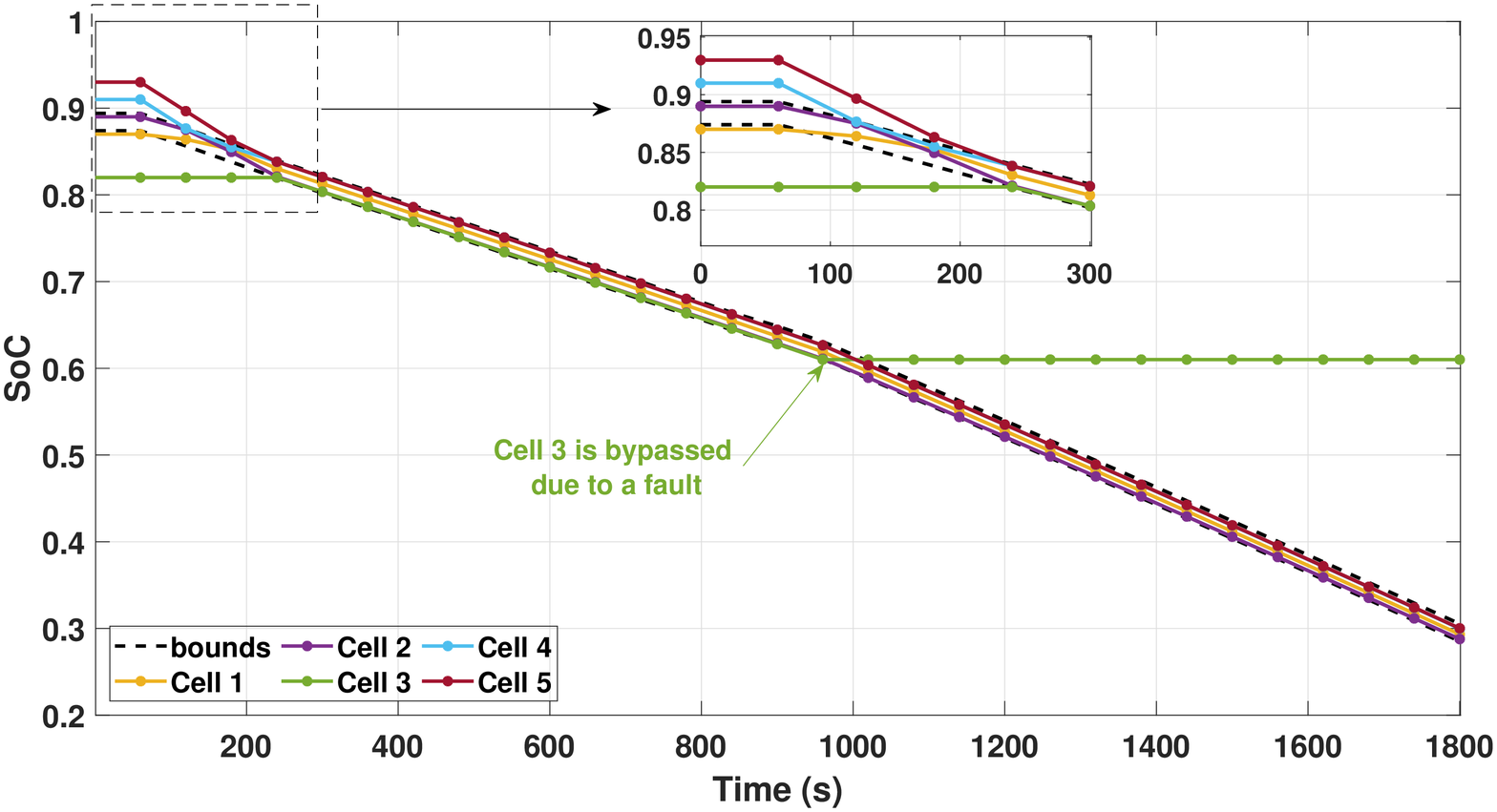} }}
	\,
    \subfloat[\centering ]{{\includegraphics[trim={2.5cm 0 2.8cm 1cm},clip,width=8.5cm]{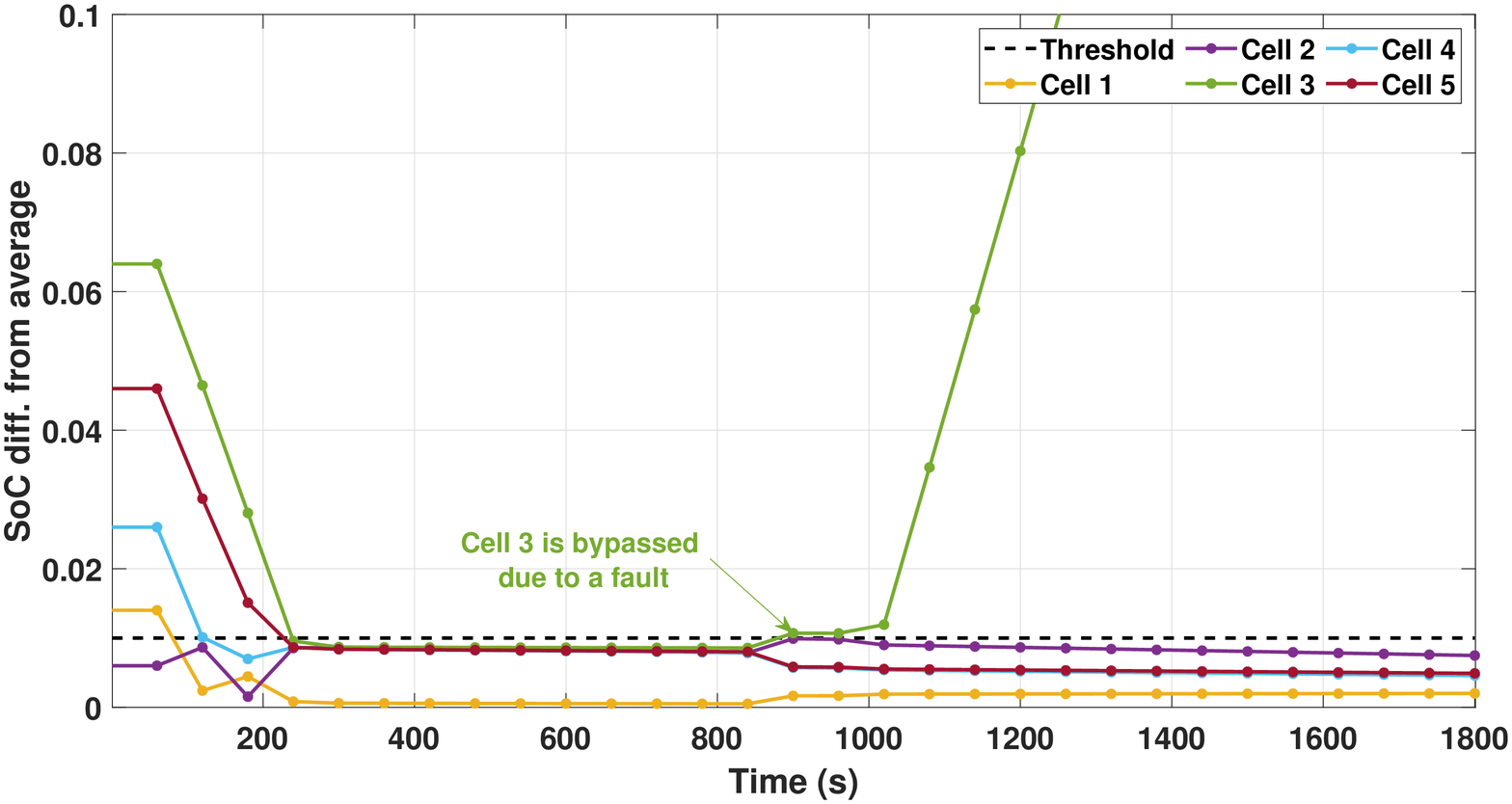} }}
    \,
    \subfloat[\centering ]{{\includegraphics[trim={2.5cm 0 2.8cm 1cm},clip,width=8.5cm]{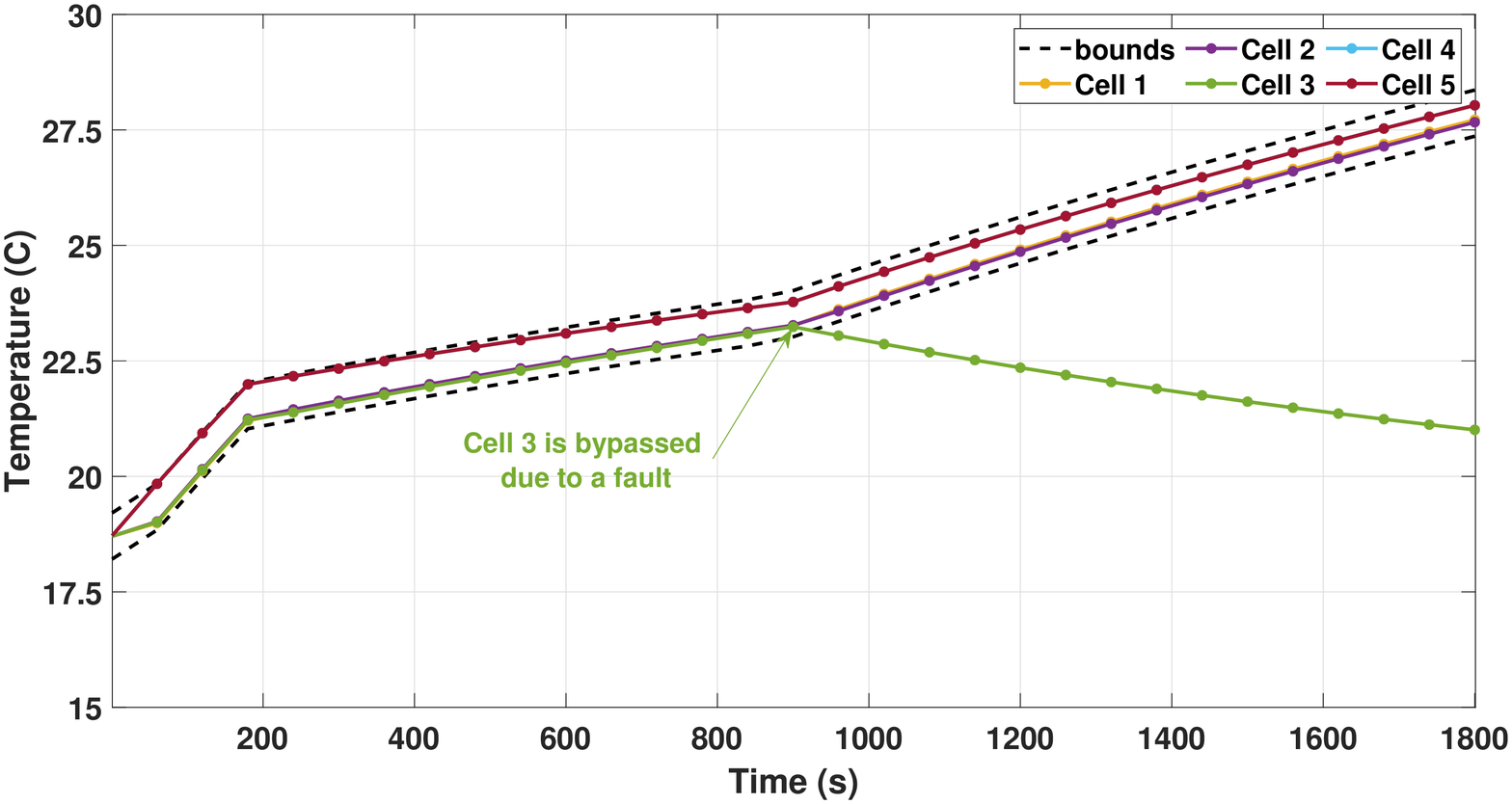} }}
    \,
    \subfloat[\centering ]{{\includegraphics[trim={2.5cm 0 2.8cm 1cm},clip,width=8.5cm]{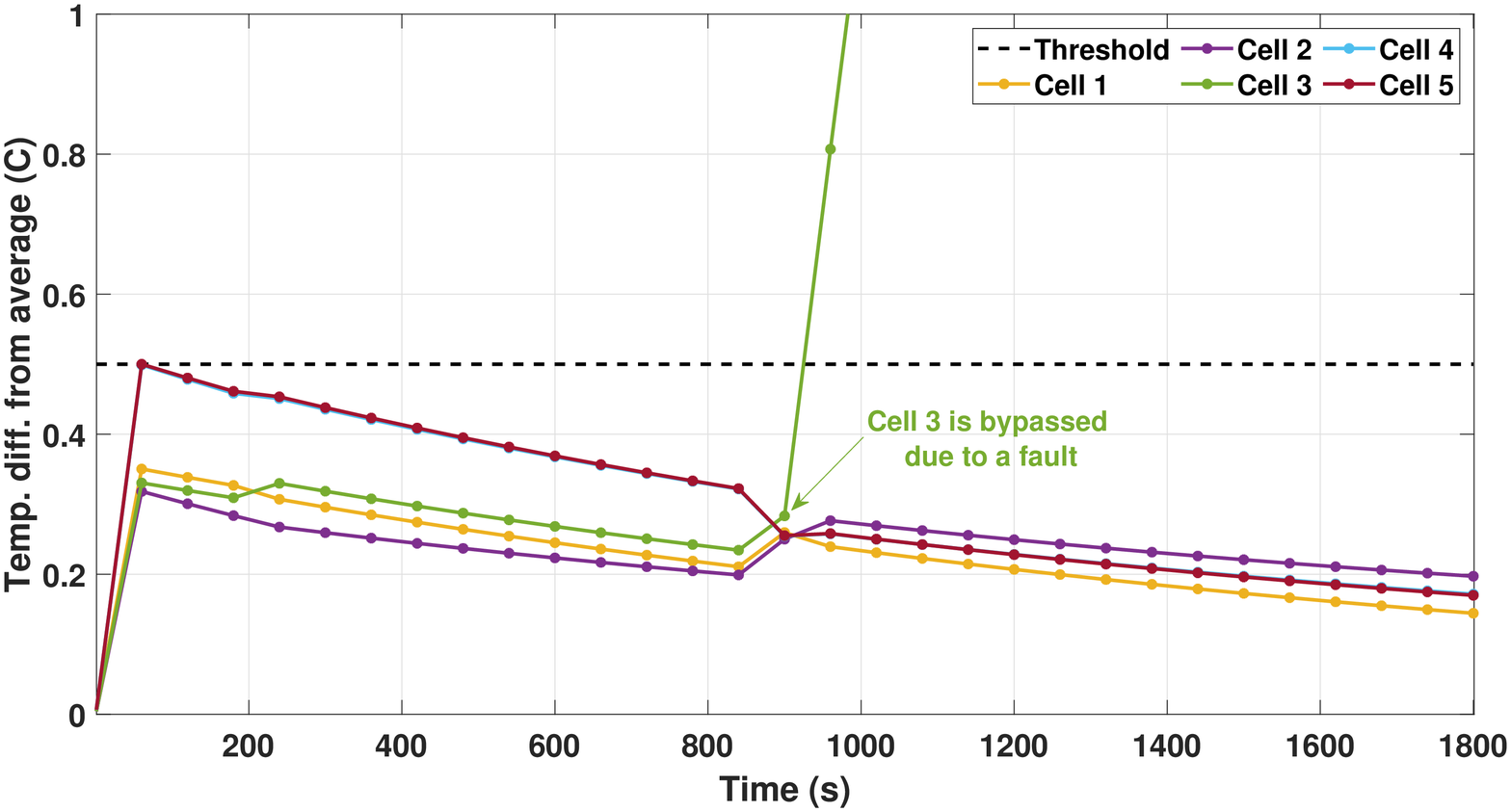} }}
    \caption{Experimental results of the proposed RBESS. (a) The SoC of the cells. (b) The temperature of the cells. (c) The SoC difference of the cells from the average. (d) The temperature difference of the cells from the average.}
    \label{FIG_10}
\end{figure*}

\begin{figure}[!t]\centering
	\includegraphics[trim={2.8cm 0cm 3cm 1cm},clip,width=\linewidth]{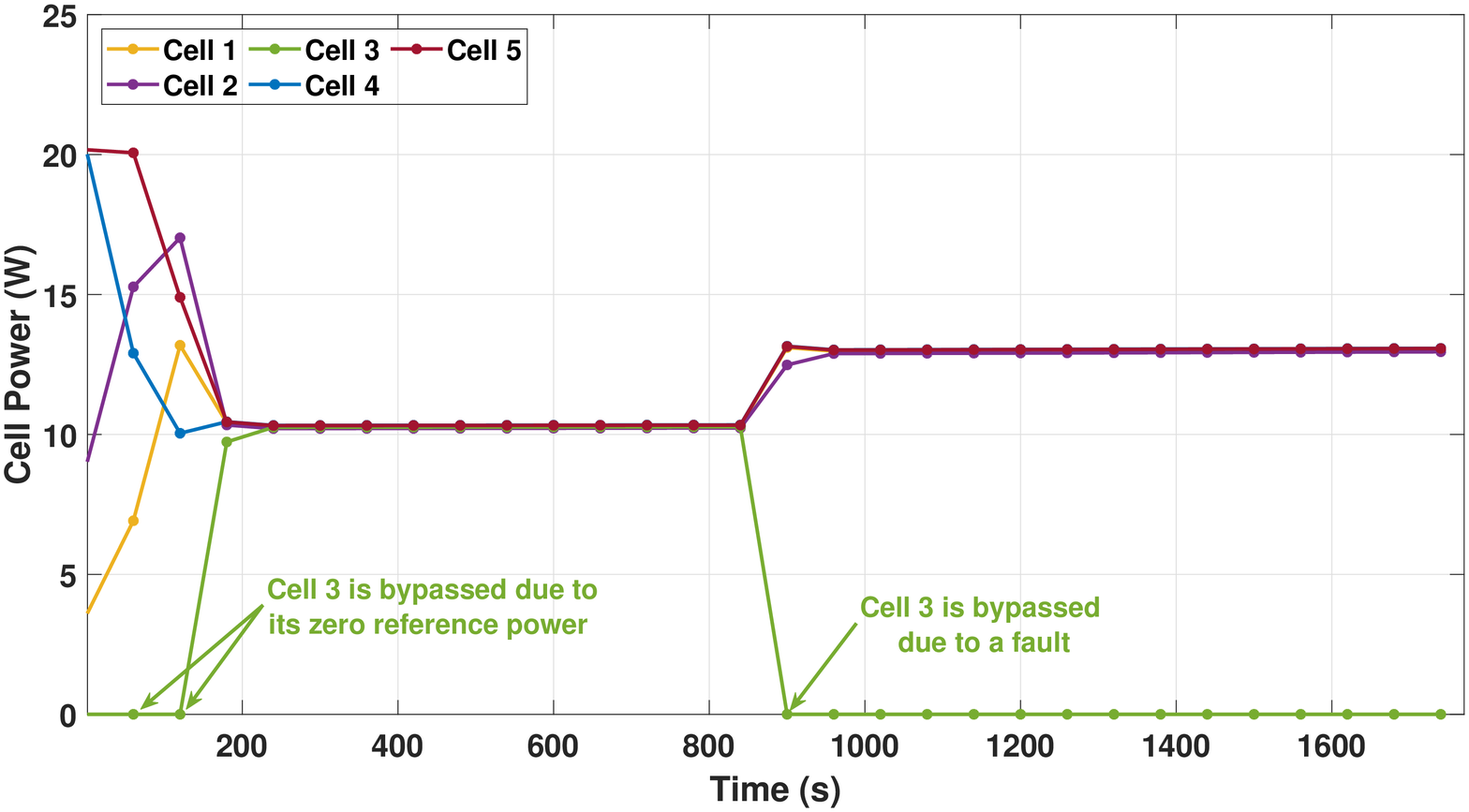}
	\caption{The output power profiles of the cells.}\label{FIG_11}
\end{figure}

The cells, labeled from 1 to 5 in order, have an initial SoC of 87\%, 89\%, 82\%, 91\%, and 93\%, respectively. The experiment lasts for 30 minutes with a sampling time of $\Delta t=60$ s. Each cell's output current is limited to 5 A. To investigate the effect of fault occurrence, a fault is assumed for cell 3 after 15 minutes of discharging in the experiment. The obtained results are shown in Figs.~\ref{FIG_10}-\ref{FIG_11}.

Fig.~\ref{FIG_10} (a) shows the SoC of the battery cells. The initial SoC values of the cells are not within the desired tolerance bound. However, the optimal power management approach successfully distributes the discharging power among the cells such that the cells reach the SoC balancing bounds after about four minutes. The corresponding output power profiles of the cells are shown in Fig.~\ref{FIG_11}. It can be seen that cell 3, with the lowest initial SoC, is assigned zero power load (and thus bypassed by reconfiguration) in the first two minutes, while cell 5, with the highest SoC, delivers the maximum allowed power. Not only does the proper distribution of the output power lead to cell balancing, but the reconfiguration capability of the proposed design also helps cell balancing. Fig.~\ref{FIG_10} (b) also shows the SoC deviation of the cells from their average value. We point out that, without the inclusion of the slack variables, the optimization would have been infeasible at the very initial moment when the SoC deviation goes beyond the SoC balancing constraint. Fig.~\ref{FIG_10} (c) depicts the temperature of the cells. The initial temperature of all the cells is 20.7$\tccentigrade$. Due to the uneven power distribution among the cells for SoC balancing, the cells will see their temperature rise and slightly drift away from each other. However, the deviation remains within the desired bound without violating the temperature balancing constraint. Fig.~\ref{FIG_10} (d) also shows the temperature deviation of the cells from the average value. The difference increases from zero to the pre-specified bound of 0.5$\tccentigrade$ in the beginning. But afterwards, it shows a declining trend and is well bounded. 

When a fault occurs to cell 3 after 15 minutes of discharging, the cell is bypassed and isolated, as indicated in Fig.~\ref{FIG_11}. Right after this happens, the other four cells that remain in service increase their discharging power accordingly, continuing to supply a total output power of 50 W as demanded. This highlights the benefit of the proposed RBESS in ensuring robust and consistent operation despite cell faults.

\section{Conclusion and Future Work}

The RBESS technology offers an important way to enhance the safe use of lithium-ion batteries. In this paper, we proposed a novel modular RBESS design, which distinguishes itself by the integration of reconfigurable power switches and DC/DC converters. The design harnesses the switching circuit reconfiguration to bypass any defective cells, and exploits the DC/DC converters to facilitate optimal power distribution at the cell level and ensure consistent power storage/supply at the system level. Based on the design, we developed a power management approach to achieve power-loss-minimized operation of the RBESS along with SoC and temperature balancing among the cells. Compared to existing methods, this approach allows wide-SoC-range operation of the cells by multi-segment SoC/OCV approximation and guarantees the feasibility of the optimization problem via mild relaxation. We conducted extensive simulations and then developed a lab-scale prototype of the RBESS design to perform validation experiments. The results substantiate the effectiveness of the proposed design and the power management approach. The study can benefit and potentially drive the use of lithium-ion batteries for safety-critical applications.

Based on the study, several interesting research questions are worth pursing further. One is how to quantify and assess the reliability of the proposed RBESS given the failure rates of the batteries, switches, and converters. A subsequent question is how to optimize the RBESS architecture and size under specified reliability metrics and power requirements. Finally, it is important to enable fast computation in large-size RBESS power management, and how to scale up the optimization approach in this paper is open to exploration.

\balance
\bibliographystyle{IEEEtran}
\scriptsize{\bibliography{IEEEabrv,Bibliography/TTE.bib}}

\begin{thebibliography}{10}
\providecommand{\url}[1]{#1}
\csname url@samestyle\endcsname
\providecommand{\newblock}{\relax}
\providecommand{\bibinfo}[2]{#2}
\providecommand{\BIBentrySTDinterwordspacing}{\spaceskip=0pt\relax}
\providecommand{\BIBentryALTinterwordstretchfactor}{4}
\providecommand{\BIBentryALTinterwordspacing}{\spaceskip=\fontdimen2\font plus
\BIBentryALTinterwordstretchfactor\fontdimen3\font minus
  \fontdimen4\font\relax}
\providecommand{\BIBforeignlanguage}[2]{{%
\expandafter\ifx\csname l@#1\endcsname\relax
\typeout{** WARNING: IEEEtran.bst: No hyphenation pattern has been}%
\typeout{** loaded for the language `#1'. Using the pattern for}%
\typeout{** the default language instead.}%
\else
\language=\csname l@#1\endcsname
\fi
#2}}
\providecommand{\BIBdecl}{\relax}
\BIBdecl

\bibitem{1512466}
A.~Affanni, A.~Bellini, G.~Franceschini, P.~Guglielmi, and C.~Tassoni,
  ``Battery choice and management for new-generation electric vehicles,''
  \emph{IEEE Transactions on Industrial Electronics}, vol.~52, no.~5, pp.
  1343--1349, 2005.

\bibitem{bills2020universal}
A.~Bills, S.~Sripad, W.~L. Fredericks, M.~Guttenberg, D.~Charles, E.~Frank, and
  V.~Viswanathan, ``Universal battery performance and degradation model for
  electric aircraft,'' \emph{arXiv preprint arXiv:2008.01527}, 2020.

\bibitem{ROSEWATER2015460}
D.~Rosewater and A.~Williams, ``Analyzing system safety in lithium-ion grid
  energy storage,'' \emph{Journal of Power Sources}, vol. 300, pp. 460--471,
  2015.

\bibitem{9353049}
F.~Stathopoulos, K.~M$\ddot{\textrm{u}}$ller, M.~Lino, T.~Kraus, P.~Klenk, and
  U.~Steinbrecher, ``Operational optimization of the lithium-ion batteries of
  \uppercase{T}erra\uppercase{SAR}-\uppercase{X}/\uppercase{T}an\uppercase{DEM}-\uppercase{X},''
  \emph{IEEE Journal of Selected Topics in Applied Earth Observations and
  Remote Sensing}, vol.~14, pp. 3243--3250, 2021.

\bibitem{FENG2018246}
X.~Feng, M.~Ouyang, X.~Liu, L.~Lu, Y.~Xia, and X.~He, ``Thermal runaway
  mechanism of lithium ion battery for electric vehicles: \uppercase{A}
  review,'' \emph{Energy Storage Materials}, vol.~10, pp. 246--267, 2018.

\bibitem{7433464}
M.~Liu, W.~Li, C.~Wang, M.~P. Polis, L.~Y. Wang, and J.~Li, ``Reliability
  evaluation of large scale battery energy storage systems,'' \emph{IEEE
  Transactions on Smart Grid}, vol.~8, no.~6, pp. 2733--2743, 2017.

\bibitem{7442763}
S.~Ci, N.~Lin, and D.~Wu, ``Reconfigurable battery techniques and systems: A
  survey,'' \emph{IEEE Access}, vol.~4, pp. 1175--1189, 2016.

\bibitem{9211456}
J.~Kim, S.~Baek, Y.~Choi, J.~Ahn, and H.~Cha, ``Hydrone: Reconfigurable energy
  storage for uav applications,'' \emph{IEEE Transactions on Computer-Aided
  Design of Integrated Circuits and Systems}, vol.~39, no.~11, pp. 3686--3697,
  2020.

\bibitem{5744772}
T.~Kim, W.~Qiao, and L.~Qu, ``Series-connected self-reconfigurable multicell
  battery,'' in \emph{Twenty-Sixth Annual IEEE Applied Power Electronics
  Conference and Exposition}, 2011, pp. 1382--1387.

\bibitem{6126056}
------, ``Power electronics-enabled self-x multicell batteries: A design toward
  smart batteries,'' \emph{IEEE Transactions on Power Electronics}, vol.~27,
  no.~11, pp. 4723--4733, 2012.

\bibitem{6165857}
S.~Ci, J.~Zhang, H.~Sharif, and M.~Alahmad, ``Dynamic reconfigurable multi-cell
  battery: A novel approach to improve battery performance,'' in
  \emph{Twenty-Seventh Annual IEEE Applied Power Electronics Conference and
  Exposition}, 2012, pp. 439--442.

\bibitem{4840570}
H.~Kim and K.~G. Shin, ``On dynamic reconfiguration of a large-scale battery
  system,'' in \emph{15th IEEE Real-Time and Embedded Technology and
  Applications Symposium}, 2009, pp. 87--96.

\bibitem{9720183}
H.~Cui, Z.~Wei, H.~He, and J.~Li, ``Novel reconfigurable topology-enabled
  hierarchical equalization of lithium-ion battery for maximum capacity
  utilization,'' \emph{IEEE Transactions on Industrial Electronics}, pp. 1--1,
  2022.

\bibitem{8099064}
A.~Singer, F.~Helling, T.~Weyh, J.~Jungbauer, and H.-J. Pfisterer, ``Modular
  multilevel parallel converter based split battery system (\uppercase{M2B})
  for stationary storage applications,'' in \emph{19th European Conference on
  Power Electronics and Applications}, 2017, pp. P.1--P.10.

\bibitem{9090983}
O.~Salari, K.~H. Zaad, A.~Bakhshai, and P.~Jain, ``Reconfigurable hybrid energy
  storage system for an electric vehicle \uppercase{DC}-\uppercase{AC}
  inverter,'' \emph{IEEE Transactions on Power Electronics}, vol.~35, no.~12,
  pp. 12\,846--12\,860, 2020.

\bibitem{7116565}
M.~Momayyezan, B.~Hredzak, and V.~G. Agelidis, ``Integrated reconfigurable
  converter topology for high-voltage battery systems,'' \emph{IEEE
  Transactions on Power Electronics}, vol.~31, no.~3, pp. 1968--1979, 2016.

\bibitem{7378996}
Y.~Li and Y.~Han, ``A module-integrated distributed battery energy storage and
  management system,'' \emph{IEEE Transactions on Power Electronics}, vol.~31,
  no.~12, pp. 8260--8270, 2016.

\bibitem{7544629}
C.~Pinto, J.~V. Barreras, E.~Schaltz, and R.~E. Ara\'{u}jo, ``Evaluation of
  advanced control for \uppercase{L}i-ion battery balancing systems using
  convex optimization,'' \emph{IEEE Transactions on Sustainable Energy},
  vol.~7, no.~4, pp. 1703--1717, 2016.

\bibitem{8768008}
R.~de~Castro, C.~Pinto, J.~Varela~Barreras, R.~E. Ara\'{u}jo, and D.~A. Howey,
  ``Smart and hybrid balancing system: Design, modeling, and experimental
  demonstration,'' \emph{IEEE Transactions on Vehicular Technology}, vol.~68,
  no.~12, pp. 11\,449--11\,461, 2019.

\bibitem{9520124}
R.~de~Castro, H.~Pereira, R.~E. Ara\'{u}jo, J.~V. Barreras, and H.~C. Pangborn,
  ``q\uppercase{TSL}: A multilayer control framework for managing capacity,
  temperature, stress, and losses in hybrid balancing systems,'' \emph{IEEE
  Transactions on Control Systems Technology}, pp. 1--16, 2021.

\bibitem{7782856}
L.~He, Z.~Yang, Y.~Gu, C.~Liu, T.~He, and K.~G. Shin,
  ``\uppercase{S}o\uppercase{H}-aware reconfiguration in battery packs,''
  \emph{IEEE Transactions on Smart Grid}, vol.~9, no.~4, pp. 3727--3735, 2018.

\bibitem{MURGOVSKI201292}
N.~Murgovski, L.~Johannesson, and J.~Sj$\ddot{\textrm{o}}$berg, ``Convex
  modeling of energy buffers in power control applications,'' \emph{IFAC
  Proceedings Volumes}, vol.~45, no.~30, pp. 92--99, 2012, 3rd IFAC Workshop on
  Engine and Powertrain Control, Simulation and Modeling.

\bibitem{7518611}
H.~Fang, Y.~Wang, and J.~Chen, ``Health-aware and user-involved battery
  charging management for electric vehicles: Linear quadratic strategies,''
  \emph{IEEE Transactions on Control Systems Technology}, vol.~25, no.~3, pp.
  911--923, 2017.

\bibitem{7995102}
Q.~Ouyang, J.~Chen, J.~Zheng, and H.~Fang, ``Optimal cell-to-cell balancing
  topology design for serially connected lithium-ion battery packs,''
  \emph{IEEE Transactions on Sustainable Energy}, vol.~9, no.~1, pp. 350--360,
  2018.

\bibitem{9589321}
A.~Farakhor and H.~Fang, ``A novel modular, reconfigurable battery energy
  storage system design,'' in \emph{47th Annual Conference of the IEEE
  Industrial Electronics Society}, 2021, pp. 1--6.

\bibitem{4012131}
S.~Chattopadhyay and S.~Das, ``A digital current-mode control technique for
  dc–dc converters,'' \emph{IEEE Transactions on Power Electronics}, vol.~21,
  no.~6, pp. 1718--1726, 2006.

\bibitem{en4040582}
H.~He, R.~Xiong, and J.~Fan, ``Evaluation of lithium-ion battery equivalent
  circuit models for state of charge estimation by an experimental approach,''
  \emph{Energies}, vol.~4, no.~4, pp. 582--598, 2011.

\bibitem{PESARAN2002377}
A.~A. Pesaran, ``Battery thermal models for hybrid vehicle simulations,''
  \emph{Journal of Power Sources}, vol. 110, no.~2, pp. 377--382, 2002.

\bibitem{TIAN2020101282}
N.~Tian, Y.~Wang, J.~Chen, and H.~Fang, ``One-shot parameter identification of
  the \uppercase{T}hevenin's model for batteries: Methods and validation,''
  \emph{Journal of Energy Storage}, vol.~29, p. 101282, 2020.

\bibitem{cvx}
M.~Grant and S.~Boyd, ``{CVX}: Matlab software for disciplined convex
  programming, version 2.1,'' \url{http://cvxr.com/cvx}, Mar. 2014.

\end{thebibliography}

\vspace{-1cm}
\begin{IEEEbiography}[{\includegraphics[width=1in,height=1.25in,clip,keepaspectratio]{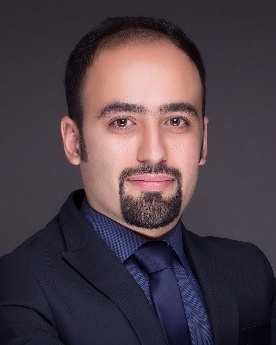}}]
{Amir Farakhor} (Student Member, IEEE) received the B.Sc. and M.Sc. degrees in electrical engineering from the Azarbaijan Shahid Madani University, Tabriz, Iran in 2012 and 2014, respectively, and Ph.D. degree in Power Electronics from the University of Tabriz in Feb 2019. He is currently a Ph.D. student in mechanical engineering at the University of Kansas, Lawrence, KS, USA. His research interests include power electronics, battery management systems, renewable energies, and distributed generation.
\end{IEEEbiography}

\vspace{-1cm}
\begin{IEEEbiography}[{\includegraphics[width=1in,height=1.25in,clip,keepaspectratio]{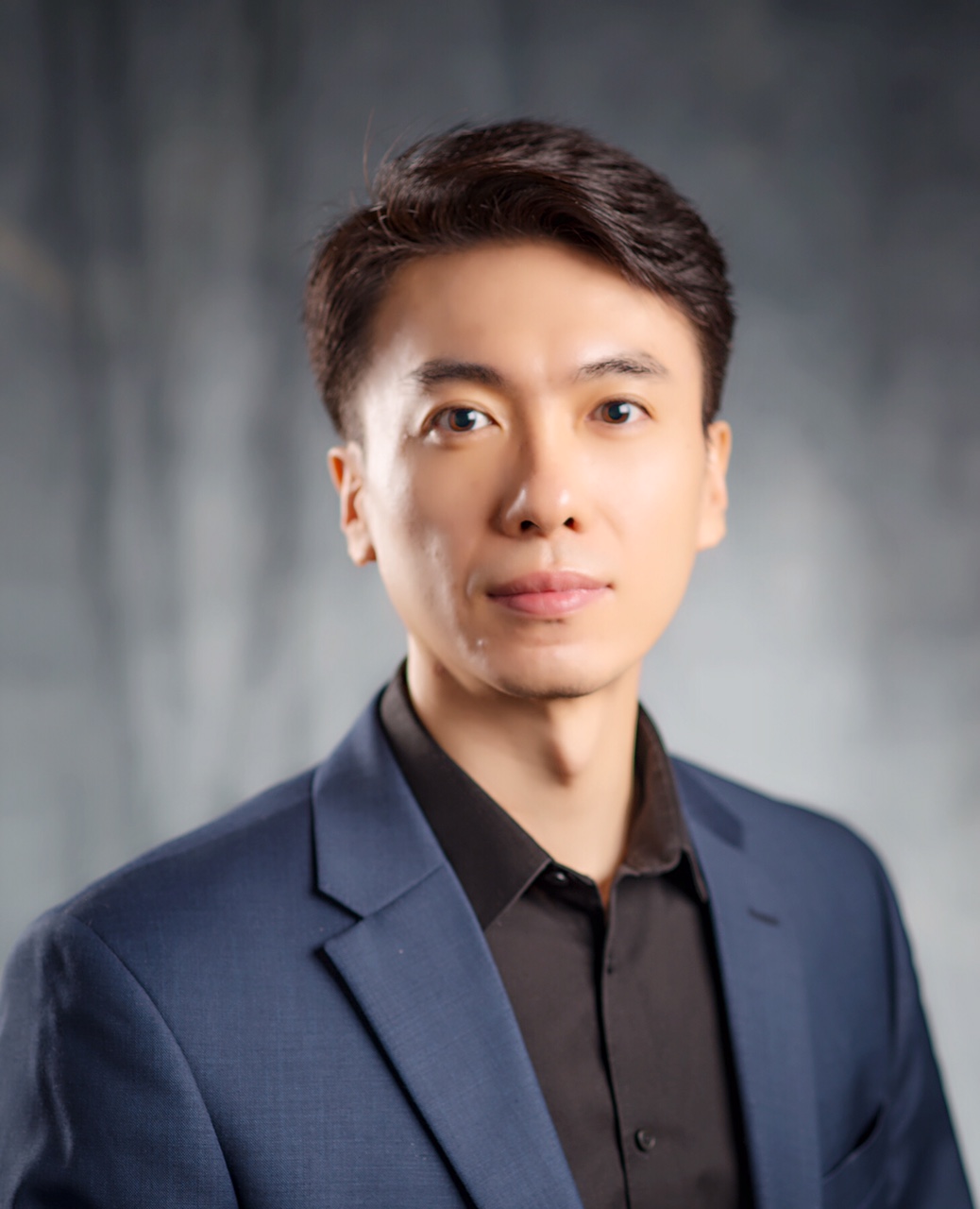}}]
{Di Wu} (Senior Member, IEEE) is a Chief Research Engineer and a Team Leader within the Optimization and Control Group at the Pacific Northwest National Laboratory (PNNL). He received the B.S. and M.S. degrees in electrical engineering from Shanghai Jiao Tong University, China, in 2003 and 2006, respectively, and the Ph.D. in electrical and computer engineering from Iowa State University, Ames, in 2012. At PNNL, Dr. Wu leads research work in the areas of energy storage analytics, building-to-grid integration, microgrid design, and hybrid energy systems. Dr. Wu is a Senior Member of IEEE and a member of the IEEE Power and Energy Society and the Control System Society. He serves as an Editor for the IEEE Open Access Journal of Power and Energy and IEEE Transactions on Energy Markets, Policy and Regulation.
\end{IEEEbiography}

\vspace{-1cm}
\begin{IEEEbiography}[{\includegraphics[width=1in,height=1.25in,clip,keepaspectratio]{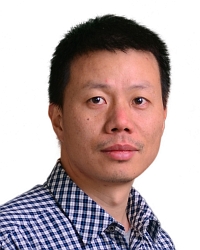}}]
{Yebin Wang} (Senior Member, IEEE) received the B.Eng. degree in mechatronics engineering
from Zhejiang University, Hangzhou, China, in 1997, the M.Eng. degree in control theory and control engineering from Tsinghua University, Beijing, China, in 2001, and the Ph.D. degree in electrical engineering from the University of Alberta, Edmonton, AB, Canada, in 2008.
He has been with Mitsubishi Electric Research Laboratories, Cambridge, MA, USA, since 2009, where he is currently a Senior Principal Research Scientist. From 2001 to 2003, he was a Software Engineer, the Project Manager, and the Manager of the Research and Development Department in Industries, Beijing, China. His current research interests include nonlinear control and estimation, optimal control, adaptive systems, and their applications, including mechatronic systems.
\end{IEEEbiography}

\vspace{-2cm}
\begin{IEEEbiography}[{\includegraphics[width=1in,height=1.25in,clip,keepaspectratio]{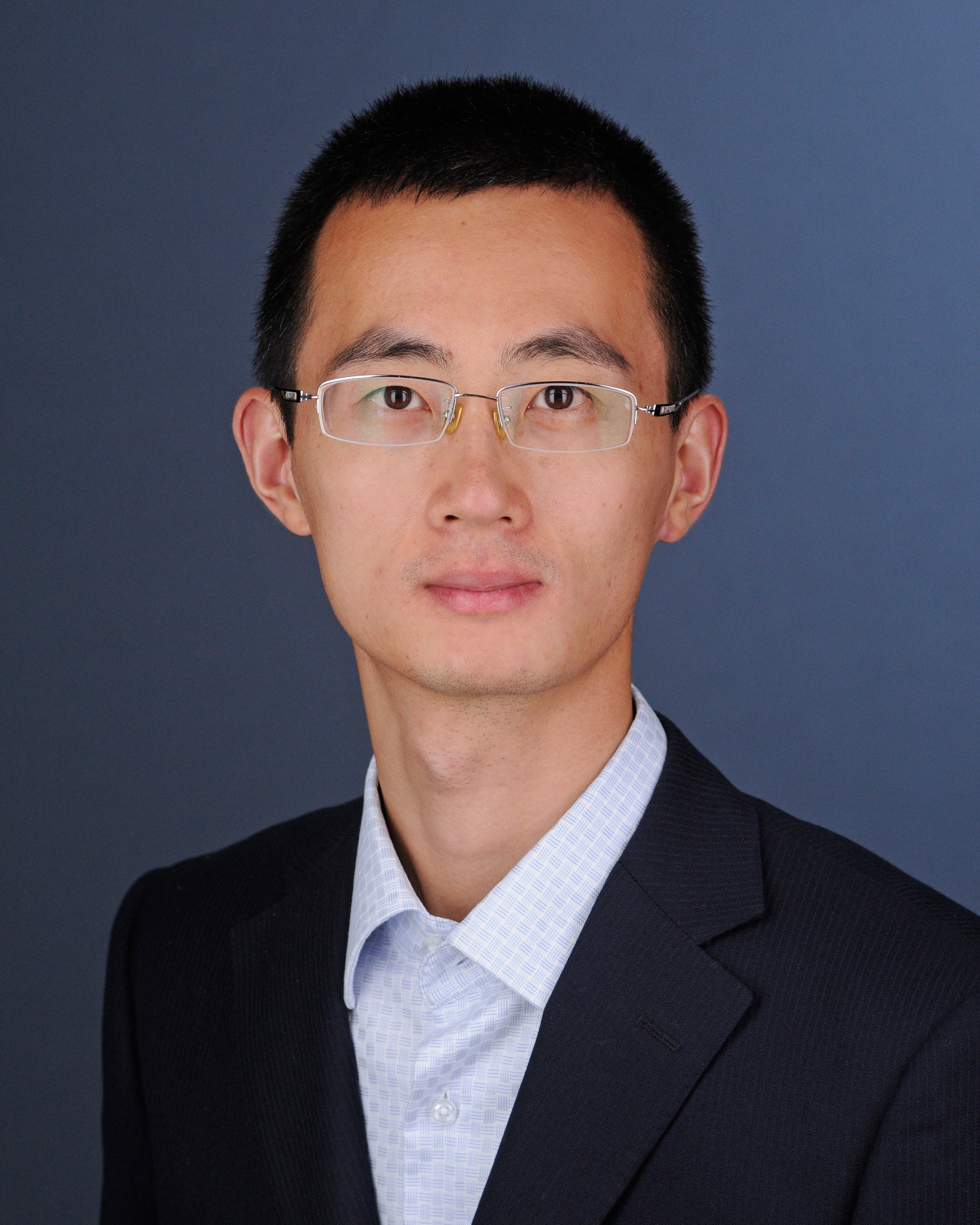}}]
{Huazhen Fang} (Member, IEEE) received the B.Eng. degree in computer science and technology from Northwestern Polytechnic University, Xi'an, China, in 2006, the M.Sc. degree in mechanical engineering from the University of Saskatchewan, Saskatoon, Canada, in 2009, and the Ph.D. degree in mechanical engineering from the Department of Mechanical and Aerospace Engineering, University of California, San Diego, La Jolla, CA, USA, in 2014. 
He is   an Associate Professor of mechanical engineering with the University of Kansas, Lawrence, KS, USA.  His research interests include control and estimation theory with application to energy management and robotics. Dr. Fang received the   National Science Foundation CAREER Award in 2019. He currently serves  as an Associate Editor for IEEE Transactions on Industrial Electronics, IEEE Open Journal of the Industrial Electronics Society, IEEE Control Systems Letters, IEEE Open Journal of Control Systems, and Information Sciences.
\end{IEEEbiography}

\end{document}